\documentclass[prb, onecolumn, notitlepage]{revtex4-1}
\usepackage{amsmath,amsfonts, amssymb, amsthm, dsfont}
\usepackage{yfonts}
\usepackage{bm}
\usepackage{mathrsfs}
\usepackage{graphicx}
\usepackage{verbatim}
\usepackage{hyperref}
\usepackage{wasysym}
\usepackage{xcolor}
\usepackage[caption=false]{subfig}

\usepackage{tikz}
\usetikzlibrary{arrows,calc,patterns,decorations.markings,shapes.geometric}
\tikzset{->-/.style={decoration={
  markings,
  mark=at position #1 with {\arrow{>}}},postaction={decorate}}
}

\newcommand{\ket}[1]{|#1\rangle}

\newcommand{\comments}[1]{}
\newcommand{\mb}[1]{\mathbf{#1}}

\newcommand{\coho}[1]{\textswab{#1}}
\newcommand{\cohosub}[1]{\scalebox{0.7}{\textswab{#1}}}

\def\U{\mathrm{U}(1)}
\def\H{\mathcal{H}}
\def\Z{\mathbb{Z}}

\def\TT{\mathsf{T}}

\makeatletter
\def\l@subsubsection#1#2{}
\makeatother

\begin{document}

\title{Exactly solvable models for U(1) symmetry-enriched topological phases}

\author{Qing-Rui Wang}
\author{Meng Cheng}
\affiliation{Department of Physics, Yale University, New Haven, CT 06511-8499, USA}
\begin{abstract}
We propose a general construction of commuting projector lattice models for 2D and 3D topological phases enriched by U(1) symmetry, with finite-dimensional Hilbert space per site. The construction starts from a commuting projector model of the topological phase and decorates U(1) charges to the state space in a consistent manner. We show that all 2D U(1) symmetry-enriched topological phases which allow gapped boundary without breaking symmetry, can be realized through our construction. We also construct a large class of 3D topological phases with U(1) symmetry fractionalized on particles or loop excitations.
\end{abstract}
\maketitle

\section{Introduction}
Exactly solvable lattice models have played a crucial role in the recent development of the theory of topological order. They provide  proofs of principle for the existence of certain topological phases, and the solubility allows a complete characterization of the universal properties of the ground state(s) as well as low-energy excitations. While exactly solubility often comes at the cost of complicated, many-body interactions, the fixed-point wavefunction illustrates the entanglement structure of the topological phase. The existence of such a fixed-point wavefunction is often a strong indication that the phase can be represented as a tensor-network state~\cite{cirac2020matrix}. In addition, the stability of topological order can be proven rigorously for commuting projector Hamiltonian models~\cite{Bravyi_2010}. 

It is therefore an important question to understand what kinds of topological phases can be realized in commuting projector Hamiltonians (CPH).
In two spatial dimensions, all topological phases with gappable boundary can be realized in generalized string-net models~\cite{Levin_2005, lin2021generalized}. More recently, the construction has been extended to topological phases enriched by a finite-group symmetry~\cite{Cheng_2017,Heinrich_2016, williamson2017symmetryenriched}, by including additional spins on the dual lattice transforming as the regular representation of $G$. Recent progress in the classification of (3+1)d topological phases also suggests that all of them have CPH realizations~\cite{Lan3D1, Lan3D2, johnsonfreyd2020classification}, including those enriched by finite-group symmetry. Besides topologically ordered phases, we note that many short-range entangled phases, such as bosonic symmetry-protected topological (SPT) phases within the group cohomology classification~\cite{Chen_2013}, can be realized by CPHs.

It is then very desirable to further generalize this construction to a continuous $G$, e.g. $G=\U$. A common approach is to represent the U(1) symmetry on quantum rotors, which are then coupled to other degrees of freedom. Along this line Ref. [\onlinecite{Motrunich_2002}] and Ref. [\onlinecite{Levin_2011}] designed exactly solvable models for insulators with fractionally charged excitations in both 2D and 3D, the topological order of which is essentially that of a $\Z_n$ lattice gauge theory.  However, it is not clear how the construction can be applied to more intricate types of topological order. Naively, the general construction in Refs. [\onlinecite{Cheng_2017}, \onlinecite{Heinrich_2016}] can be generalized to U(1) symmetry represented by rotors, but then one runs into issues of having Hamiltonian terms being discontinuous functions of the angular variable, and it is not entirely clear how to make sense of the Hamiltonian. Related rotor models for fractionalized insulators were also studied in Refs. [\onlinecite{Geraedts_2013}, \onlinecite{Chen_2019}]. As we have noted, all of the existing models exploit rotors as natural state space for U(1) charges.

If we further restrict to models with finite-dimensional Hilbert space per site, much fewer examples are known even for SPT phases. Quite recently, Refs. [\onlinecite{metlitski20191d},\onlinecite{Son_2019}] presented CPH realizations of 2D electronic time-reversal-invariant topological insulators (TI), and Ref. [\onlinecite{horinouchi2020solvable}] constructed a CPH for 2D bosonic topological insulator.  In this direction, another important advance is a no-go theorem proved in Ref. [\onlinecite{Kapustin_2019}], showing that CPHs with finite-dimensional site Hilbert space must have zero Hall conductance.

In this work, we provide a systematic construction of CPH models for a broad class of topological phases with U(1) symmetry. Generalizing the idea in Ref. [\onlinecite{horinouchi2020solvable}], the basic strategy is to start from a CPH model and decorate the state space with U(1) charges. Our construction includes \emph{any} 2D non-chiral topological order enriched by U(1), as long as the Hall conductance vanishes. In some sense, this is the best one can hope for in view of the theorem in Ref. [\onlinecite{Kapustin_2019}].  We also show that similar ideas can be used to construct a large class of 3D topological order with U(1) symmetry. In particular, we describe how to exactly realize \emph{any} 2D topological phase with U(1) symmetry on the surface of a 3D CPH of Walker-Wang type~\cite{walker201131tqfts}, with no restriction on the topological order or the Hall conductance. With this construction, we describe a CPH for the bosonic topological insulator.  We stress that our construction is done with only finite-dimensional local Hilbert space, compared to many existing models mentioned earlier.

\section{General construction}
\label{sec:general}
We first sketch the idea of the general construction. 

All the models discussed in this work are derived from the so-called ``state-sum" construction of topological quantum field theories. These models can be defined on any spatial manifold, as long as a simplicial triangulation is provided. In addition, we also require a ``branching structure", referring to orientations on all simplicies induced by an ordering of vertices (0-simplicies). The physical degrees of freedom are placed on various simplices, e.g. vertices, links, or faces. In other words, each simplex is associated with a finite-dimensional Hilbert space. Typically there is a natural basis for the local Hilbert space, which can be related to a certain algebraic structure. We refer to the basis states of the whole system as configurations.

The Hamiltonian consists of two types of terms. The first kind of terms imposes certain local constraints to select a subset of the configurations as the low-energy subspace. The second kind of terms then introduces local ``moves" between low-energy configurations, which can be interpreted as relations among amplitudes of configurations that only differ from each other in a small local patch.  The ground state is a superposition of all allowed configurations, with amplitudes determined uniquely by the local moves. Schematically we can write
\begin{equation}
    \ket{\Psi_0}=\sum_{C}\Psi(C)\ket{C}.
\end{equation}
Here $C$ stands for allowed configurations.

To incorporate U(1) symmetry, we introduce spins to the top simplices (e.g. faces in two dimensions). In the examples below, it is sufficient to consider spin-$1$, where the U(1) charge is $Q=S^z\in\{-1,0,1\}$. For each allowed configuration, the charge $S^z$ values are uniquely fixed by the local configuration state of the simplex. Thus we say that the configurations are ``decorated" by U(1) charges. The charge decoration must be compatible with local moves. Namely, any two configurations related by local moves have to have the same charge. The ground state wavefunction becomes
\begin{equation}
    \ket{\Psi_0}=\sum_{C}\Psi(C)\ket{C; \{Q\}}.
\end{equation}
Here $\ket{C;\{Q\}}$ denote configurations with charges decorated. Note that crucially the amplitudes stay the same. It is not difficult to modify the Hamiltonian: the local constraints now also enforce the charge decoration rules, and the local moves only operate within the configurations with the right decorations.

In the following, we will give more details of the construction in both 2D and 3D.

\section{$\Z_2$ topological order in two dimensions}
We first demonstrate the construction for the $\Z_2$ toric code topological order, realized on a honeycomb lattice. On each edge, we place a qubit, and we think of the two basis states as the edge being occupied by a ``string" or not.
We impose the constraint that strings must form closed loops, which means that there must be an even number of strings meeting at every vertex. The ground state is an equal-weight superposition of all closed string states. There are two kinds of excitations: the $e$ excitation is where an odd number of strings terminating at a vertex. The $m$ excitation, centered at a hexagon, does not violate any vertex constraints. They are defined by introducing additional signs in the amplitudes of string states according to winding numbers of strings around the excitations. It can be shown that the $e$ and $m$ excitations can be pair created out of the ground state. Both $e$ and $m$ are bosons but $e$ and $m$ have a mutual braiding statistics $-1$.

We now place an additional spin-$1$ degree of freedom on each site. The total charge of the system is given by
\begin{equation}\label{Q}
    Q=\sum_v S^z_v.
\end{equation}
For each loop configuration in the ground states, the spin $S^z_v$ are fixed by its three edges, according to the following rules:
\begin{equation}\label{fig:Z2}
\vcenter{\hbox{
\begin{tikzpicture}[scale=0.25,>=stealth]
    \def\shift{0};
    \def\shifty{0}
    \draw[blue, very thick] (\shift+0,\shifty+4)  -- (\shift+2,\shifty+2) ;
    \draw[blue, very thick]  (\shift+4,\shifty+4) -- (\shift+2,\shifty+2) ;
    \draw[very thick,dotted](\shift+2,\shifty+2) -- (\shift+2,\shifty+0);
    \filldraw (\shift+2, \shifty+2.1) circle (0.3) node[above]{$+$};
    
    \def\shift{8};
    \def\shifty{0}
    \draw[dotted, very thick] (\shift+0,\shifty+4) -- (\shift+2,\shifty+2) ;
    \draw[blue, very thick]  (\shift+4,\shifty+4) -- (\shift+2,\shifty+2) ;
    \draw[blue, very thick](\shift+2,\shifty+2) -- (\shift+2,\shifty+0) ;
    
    \def\shift{16};
    \def\shifty{0}
    \draw[blue, very thick] (\shift+0,\shifty+4) -- (\shift+2,\shifty+2) ;
    \draw[dotted, very thick]  (\shift+4,\shifty+4) -- (\shift+2,\shifty+2) ;
    \draw[blue, very thick](\shift+2,\shifty+2) -- (\shift+2,\shifty+0) ;
    
    \def\shift{24};
    \def\shifty{0}
    \draw[blue, very thick] (\shift+0,\shifty+0)  -- (\shift+2,\shifty+2) ;
    \draw[blue, very thick]  (\shift+4,\shifty+0) -- (\shift+2,\shifty+2) ;
    \draw[very thick,dotted](\shift+2,\shifty+2) -- (\shift+2,\shifty+4);
    \filldraw[black] (\shift+2, \shifty+1.9) circle (0.3) node[below]{$-$};
    \end{tikzpicture}
}}
\end{equation}
The ground state is a superposition of all closed-loop configurations with the same amplitudes.

The exactly solvable commuting projector Hamiltonian of the model reads
\begin{equation}\label{H}
    H=-\sum_v A_v - \sum_p B_p.
\end{equation}
The vertex term $A_v$ enforces both the closed-string rule and the $\U$ charge decoration rule shown in Eq.~(\ref{fig:Z2}). The plaquette term $B_p$ is designed to fluctuate the loop configurations with the $\U$ charge conserved. The details of the Hamiltonian are given in Appendix~\ref{App:H}. We note that the charge decoration breaks the $C_3$ rotation as well as the sublattice symmetry of the underlying honeycomb lattice.


In the following, we present a direct calculation of the fractional charge carried by an $e$ excitation. First, we explain how the fractional charge of excitation is defined in a general gapped phase. For simplicity, we assume that the system is translation-invariant. Suppose an excitation $a$ is localized at position $\mb{R}$. Let $B_l$ be a disk of radius $l$ centered at $\mb{R}$, assuming that no other excitations are found in the disk. We define
\begin{equation}
    Q_{\mb{R}}(l) = \sum_{\mb{r}\in B_l}\Delta Q_\mb{r},\quad \Delta Q_\mb{r}=\langle Q_\mb{r}\rangle-\langle Q_\mb{r}\rangle_0,
\end{equation}
where $\langle \cdot \rangle$ is the expectation value over the state with the excitation, and $\langle \cdot \rangle_0$ is the ground state expectation value. $Q_\mb{r}$ is the charge operator at site $\mb{r}$. The fractional charge carried by the excitation is then
\begin{equation}
    Q_a=\lim_{l\rightarrow \infty} Q_\mb{R}(l) \text{ mod }1.
\end{equation}
One caveat in this definition is that as $l$ goes to infinity, all the other excitations must be kept far away from $B_l$. Because of the energy gap, $\langle Q_\mb{r}\rangle-\langle Q_\mb{r}\rangle_0$ decays exponentially away from the location of the excitation, and thus $Q_a$ is well-defined. For a fixed-point model with zero correlation length, we expect that the distribution of $\Delta Q_\mb{r}$ is strictly short-range without any exponential tail, so the limiting procedure is not really necessary.

We apply the definition to the model Eq.~\eqref{H}. First we calculate the average charge density in the ground state. Note that the ground state breaks the sublattice symmetry of the honeycomb lattice, as the sublattice $A$ ($B$) in the first (last) figure of Eq.~(\ref{fig:Z2}) carries $\U$ charge $+1$ ($-1$). For a given vertex $v$ of, say, sublattice $A$, there are 4 possible closed string configurations of its three connecting edges in the ground state:
\begin{align}\label{Psi_0}
    |\Psi_0\rangle
    &\sim
    \Bigg|
    \vcenter{\hbox{\!
    \begin{tikzpicture}[scale=0.2,>=stealth]
    \def\shift{0};
    \def\shifty{0}
    \draw[dotted, very thick] (\shift+0,\shifty+4)  -- (\shift+2,\shifty+2) ;
    \draw[dotted, very thick]  (\shift+4,\shifty+4) -- (\shift+2,\shifty+2) ;
    \draw[very thick,dotted](\shift+2,\shifty+2) -- (\shift+2,\shifty+0);
    \end{tikzpicture}
    }}
    \!\Bigg\rangle
    +
    \Bigg|
    \vcenter{\hbox{\!
    \begin{tikzpicture}[scale=0.2,>=stealth]
    \def\shift{0};
    \def\shifty{0}
    \draw[blue, very thick] (\shift+0,\shifty+4)  -- (\shift+2,\shifty+2) ;
    \draw[blue, very thick]  (\shift+4,\shifty+4) -- (\shift+2,\shifty+2) ;
    \draw[very thick,dotted](\shift+2,\shifty+2) -- (\shift+2,\shifty+0);
    \filldraw (\shift+2, \shifty+2.1) circle (0.3) node[above]{$+$};
    \end{tikzpicture}
    }}
    \!\Bigg\rangle
    +
    \Bigg|
    \vcenter{\hbox{\!
    \begin{tikzpicture}[scale=0.2,>=stealth]
    \def\shift{0};
    \def\shifty{0}
    \draw[dotted, very thick] (\shift+0,\shifty+4)  -- (\shift+2,\shifty+2) ;
    \draw[blue, very thick]  (\shift+4,\shifty+4) -- (\shift+2,\shifty+2) ;
    \draw[very thick,blue](\shift+2,\shifty+2) -- (\shift+2,\shifty+0);
    \end{tikzpicture}
    }}
    \!\Bigg\rangle
    +
    \Bigg|
    \vcenter{\hbox{\!
    \begin{tikzpicture}[scale=0.2,>=stealth]
    \def\shift{0};
    \def\shifty{0}
    \draw[blue, very thick] (\shift+0,\shifty+4)  -- (\shift+2,\shifty+2) ;
    \draw[dotted, very thick]  (\shift+4,\shifty+4) -- (\shift+2,\shifty+2) ;
    \draw[very thick,blue](\shift+2,\shifty+2) -- (\shift+2,\shifty+0);
    \end{tikzpicture}
    }}
    \!\Bigg\rangle.
\end{align}
Among the 4 configurations, only one of them has a nonzero charge $Q_v=S_v^z=\pm 1$ for sublattice $A/B$. Therefore, the vertex $v$ has background vacuum charge expectation value $\langle Q_v\rangle_0=\pm 1/4$. For a lattice with an integral number of unit cells, the total $\U$ charge of the ground state is still zero. 

Let us now demonstrate that $e$ anyons carry half charges (relative to the background vacuum charge). The $e$ anyons are created in pairs as endpoints of an open string operator acting on the ground state. To make the excited state an eigenstate of all the plaquette terms, we also need to apply a projection operator $\prod_p B_p =\prod_p \frac{1+B_p^s}{2}$, where $B_p^s$ create a loop $s$ around the plaquette $p$. It makes the shape of the open string fluctuate while the endpoints fixed. 
For the vertices other than the endpoints of the string operator, the average $\U$ charges in the excited states are the same as the ground state. 
To measure the fractional charge at exactly the endpoint $v$, we only need to consider the three nearby plaquette operators which may change the charge $Q_v$ at $v$ (other plaquette operators $B_p$ commute with $Q_v$). 
As an example, let us assume that $v$ is the upper endpoint of an open string in the initial state. By construction, we choose the open string operator such that $Q_v=-1$ in the initial state (the final fractional charge of $v$ does not depend on this choice). The local patch of the excited state would look like
\begin{align}\label{Psi_e}
    |\Psi_e\rangle &\sim
    B_{p_3}B_{p_2}B_{p_1}
    \Bigg|
    \vcenter{\hbox{\!\!
    \begin{tikzpicture}[scale=0.2,>=stealth]
    \def\shift{0};
    \def\shifty{0}
    \draw[dotted, very thick] (\shift+0,\shifty+4)  -- (\shift+2,\shifty+2) ;
    \draw[dotted, very thick]  (\shift+4,\shifty+4) -- (\shift+2,\shifty+2) ;
    \draw[very thick,blue](\shift+2,\shifty+2) -- (\shift+2,\shifty+0);
    \filldraw (\shift+2, \shifty+2.1) circle (0.3) node[above,yshift=-1.5]{$-$};
    \node[] at (2,4){$p_1$};
    \node[] at (.4,1.4){$p_3$};
    \node[] at (3.6,1.4){$p_2$};
    \end{tikzpicture}
    }}
    \!\!\Bigg\rangle\\\nonumber
    &\sim
    \Bigg|
    \vcenter{\hbox{\!
    \begin{tikzpicture}[scale=0.2,>=stealth]
    \def\shift{0};
    \def\shifty{0}
    \draw[dotted, very thick] (\shift+0,\shifty+4)  -- (\shift+2,\shifty+2) ;
    \draw[dotted, very thick]  (\shift+4,\shifty+4) -- (\shift+2,\shifty+2) ;
    \draw[very thick,blue](\shift+2,\shifty+2) -- (\shift+2,\shifty+0);
    \filldraw (\shift+2, \shifty+2.1) circle (0.3) node[above]{$-$};
    \end{tikzpicture}
    }}
    \!\Bigg\rangle
    +
    \Bigg|
    \vcenter{\hbox{\!
    \begin{tikzpicture}[scale=0.2,>=stealth]
    \def\shift{0};
    \def\shifty{0}
    \draw[blue, very thick] (\shift+0,\shifty+4)  -- (\shift+2,\shifty+2) ;
    \draw[dotted, very thick]  (\shift+4,\shifty+4) -- (\shift+2,\shifty+2) ;
    \draw[very thick,dotted](\shift+2,\shifty+2) -- (\shift+2,\shifty+0);
    \end{tikzpicture}
    }}
    \!\Bigg\rangle
    +
    \Bigg|
    \vcenter{\hbox{\!
    \begin{tikzpicture}[scale=0.2,>=stealth]
    \def\shift{0};
    \def\shifty{0}
    \draw[dotted, very thick] (\shift+0,\shifty+4)  -- (\shift+2,\shifty+2) ;
    \draw[blue, very thick]  (\shift+4,\shifty+4) -- (\shift+2,\shifty+2) ;
    \draw[very thick,dotted](\shift+2,\shifty+2) -- (\shift+2,\shifty+0);
    \end{tikzpicture}
    }}
    \!\Bigg\rangle
    +
    \Bigg|
    \vcenter{\hbox{\!
    \begin{tikzpicture}[scale=0.2,>=stealth]
    \def\shift{0};
    \def\shifty{0}
    \draw[blue, very thick] (\shift+0,\shifty+4)  -- (\shift+2,\shifty+2) ;
    \draw[blue, very thick]  (\shift+4,\shifty+4) -- (\shift+2,\shifty+2) ;
    \draw[very thick,blue](\shift+2,\shifty+2) -- (\shift+2,\shifty+0);
    \end{tikzpicture}
    }}
    \!\Bigg\rangle.
\end{align}
One explicit example of the string configurations for a larger patch of the lattice is shown in Appendix~\ref{App:frac}. If we act on Eq.~(\ref{Psi_e}) by more plaquette operators $B_p$ far away from the vertex $v$, each of the four terms in the second line will split into more terms with $Q_v$ unchanged. This is why we only need to consider three plaquette operators to measure $Q_v$. 
Among the four string configurations in Eq.~(\ref{Psi_e}), the last three must have zero charge at the vertex $v$. This is simply because there is a maximum point with $-1$ charge for the string to go down (see Appendix~\ref{App:frac}). Therefore, the expectation value of the $\U$ charge of $v$ is $-1/4$. And the charge of $v$ with respect to the ground state background is
\begin{equation}
    \langle Q_v\rangle - \langle Q_v\rangle_0 = -1/4 - 1/4 = -1/2.
\end{equation}
This is exactly the desired half charge for $e$ anyons of the toric code. We see that the fractional charge is highly localized at the string endpoint in this fixed-point model with zero correlation length.

Another way to show fractional charge of $e$ is to use the notion of cluster charge introduced in Ref.~\onlinecite{Levin_2011}. In our model, we define the cluster charge of vertex $v$ to be
\begin{equation}\label{Qcluster}
\begin{aligned}
    Q_v^\mathrm{cluster}
    &= 2 Q_v \mp ([a_v]+[b_v]-[c_v]).
\end{aligned}
\end{equation}
The $\mp$ signs depend on the sublattice $A/B$ of $v$. 
It can be shown that $Q_v^\mathrm{cluster}$ commutes with both $A_v$ and $B_p$ terms of the Hamiltonian, so it is a conserved quantity. It is zero in the ground state configurations given in Eq.~\eqref{fig:Z2}, and an odd integer in the presence of $e$ anyon ($[a_v+b_v-c_v]=1$).
For any region $S$ of the lattice, the total cluster charge is
\begin{equation}\label{sumQv}
    \sum_{v\in S} Q_v^\mathrm{cluster}
    = 2\sum_{v\in S} Q_v + \sum_{v\in S,v'\in \bar S} \pm a_{vv'},
\end{equation}
where $a_{vv'}$ is the string label of edge connecting vertices $v$ and $v'$. If we consider the cluster charge difference of an excited state with an $e$ anyon inside $S$ and the ground state without $e$ anyons, we have
\begin{equation}
    \left\langle \sum_{v\in S} Q_v \right\rangle - \left\langle \sum_{v\in S} Q_v \right\rangle_0
    =
    \frac{1}{2}
    \left( \left\langle \sum_{v\in S} Q_v^\mathrm{cluster} \right\rangle - \left\langle \sum_{v\in S} Q_v^\mathrm{cluster} \right\rangle_0 \right)
    =\frac{1}{2}\quad (\text{mod }1).
\end{equation}
The contribution from the second term in Eq.~\eqref{sumQv} vanishes, as it is a boundary term far from the $e$ anyon when $S$ is large enough. In this way, we again obtained the half fractional charge carried by $e$ the anyon.

For the $m$ anyon, since they do not involve any vertex violations, $m$ carries no $\U$ charge. 

We note that the construction can be effortlessly adopted to a similar model of double semion topological order~\cite{Levin_2005}. The only difference between the two models is that the wavefunction of the double semion has an additional sign $(-1)^{N_c}$, where $N_c$ is the number of loops in the configuration, but this sign factor does not interfere with the charge decorations at all. As a result, all our results carry over straightforwardly, and we conclude that an excitation violating the vertex constraint carries a half charge. Such excitations are identified as the semion or the anti-semion in the double semion topological order~\cite{Levin_2005}.  

\section{$\Z_n$ topological order}
\label{sec:Zn}
We now generalize the construction to the $\Z_n$ toric code. Now each edge has a $\Z_n$ spin, whose orthonormal basis is labeled by $0,1,\dots,n-1$ mod $n$. We again decorate each vertex by spin-$1$, and the $\U$ charge of vertex $v$ is $Q_v=S_v^z$. The following vertex constraints are imposed:
\begin{equation}\label{ZnV}
\vcenter{\hbox{
\begin{tikzpicture}[scale=0.25,>=stealth]
    \def\shift{0};
    \def\shifty{0}
	\draw[very thick] (\shift+0,\shifty+4) node[above]{$a$}  -- (\shift+2,\shifty+2) ;
	\draw[ very thick]  (\shift+4,\shifty+4) node[above]{$b$} -- (\shift+2,\shifty+2) ;
	\draw[ very thick](\shift+2,\shifty+2) -- (\shift+2,\shifty+0) node[below]{$c=[a+b]$};
    \filldraw (\shift+2, \shifty+2.1) circle (0.3) ;
	\node at (\shift+2, \shifty+3.5) {$+q$};
    
    \def\shift{8};
    \def\shifty{0}
	\draw[ very thick] (\shift+0,\shifty+0) node[below]{$a$}  -- (\shift+2,\shifty+2) ;
	\draw[ very thick]  (\shift+4,\shifty+0) node[below]{$b$} -- (\shift+2,\shifty+2) ;
	\draw[ very thick](\shift+2,\shifty+2) -- (\shift+2,\shifty+4) node[above]{$c=[a+b]$};
    \filldraw (\shift+2, \shifty+1.9) circle (0.3);
    \node at (\shift+2, \shifty+0.5) {$-q$};
    \end{tikzpicture}
}},
\end{equation}
where $q$ depends on the edge labels as
\begin{equation}\label{q}
    q(a,b)
    =\frac{1}{n}([a]+[b]-[a+b]).
\end{equation}
Here, $[a]$ is defined as $a$ (mod $n$). Mathematically, the expression of $q(a,b)$ is the generating 2-cocycle in $\H^2[\Z_n,\Z]=\Z_n$.

Again, the exactly solvable CPH consists of two kinds of terms $A_v$ and $B_p$ as in Eq.~(\ref{H}). The details of the Hamiltonian are given in Appendix~\ref{App:H}. The vertex term $A_v$ enforces the condition $[c]=[a+b]$ and the charge decorations shown in Eqs.~(\ref{ZnV}) and (\ref{q}). The plaquette term $B_p$ is a summation
\begin{equation}
    B_p=\frac{1}{n}\sum_{g\in\Z_n}B_p^{(g)},
\end{equation}
where $B_p^{(g)}$ creates a loop labeled by $g\in \Z_n$ around the plaquette $p$ and then fuses it with the six edges. The ground state is an equal-weight superposition of all allowed string configurations with $\U$ charge decorations.

For a vertex $v$ of sublattice $A$ or $B$ in Eq.~(\ref{ZnV}), we can calculate the $\U$ charge in the ground state by averaging over all allowed string configurations:
\begin{equation}\label{Qv0Zn}
    \langle Q_v\rangle_0
    =\frac{1}{n^2}\sum_{a,b,c\in\Z_n}\pm q(a,b)\delta_{[c],[a+b]}
    = \pm\frac{n-1}{2n}.
\end{equation}
It is a simple generalization of the vacuum charge $\pm 1/4$ for $n=2$ discussed in the previous section.

The $e$ anyons of the $\Z_n$ topological order can be also shown to carry $1/n$ (mod 1) fractional $\U$ charges (with respect to the ground state background charge). Similar to the $n=2$ case, $e$ anyons are created by pair as endpoints of an open string operator, followed by a projection operator $\prod_p B_p$ to fluctuate the strings. For vertices different from the string endpoints, the local string configurations are indistinguishable from the ground state. So the average charges are the same as Eq.~(\ref{Qv0Zn}) of the ground state. 
Now let us calculate the fractional charge of $e$ anyons at the string endpoints. Suppose $v$ is the upper endpoint of a string labeled by $\alpha\in\Z_n$ ($\alpha\ne 0$) with charge $Q_v=-1$ as shown in the left figure of Eq.~(\ref{B3}). Since all the $B_p$ operators away from $v$ commute with charge $Q_v$, we only need to consider $B_p$ of the three nearby plaquettes. After the action of $B_{p_3}^{(g_3)}B_{p_2}^{(g_2)}B_{p_1}^{(g_1)}$, the string configuration becomes
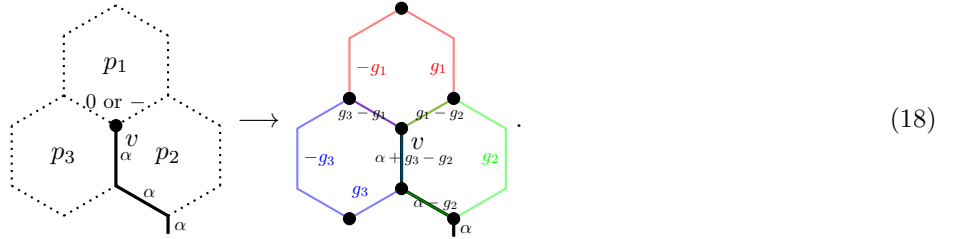
\begin{align}\label{B3}
\vcenter{\hbox{
\begin{tikzpicture}[scale=0.8,>=stealth]
    \def\p1{1}
    \coordinate (p1) at (0,1);
    \coordinate (p2) at (0.8660,-.5);
    \coordinate (p3) at (-0.8660,-.5);
    \node[below right] at (0,0) {$v$};
    \foreach \p in {1,2,3} {
        \node[]at (p\p) {$p_\p$};
        \foreach \i in {0,1,2,3,4,5} \coordinate (p\p\i) at ($(p\p)+(90-60*\i:1)$);
    }
    \foreach\i in {0,1,2,3,4,5} \draw[dotted,thick]($(p1)+(90-60*\i:1)$)--($(p1)+(90-60*\i+60:1)$);
    \foreach\i in {1,2,3,4,5} \draw[dotted,thick]($(p2)+(90-60*\i:1)$)--($(p2)+(90-60*\i+60:1)$);
    \foreach\i in {3,4,5,6} \draw[dotted,thick]($(p3)+(90-60*\i:1)$)--($(p3)+(90-60*\i+60:1)$);
    \draw[black,very thick](p23)--($(p23)+(0,-.3)$);
    \draw[black,very thick](p23)--(p24)--(p25);
    \filldraw (0,0) circle (0.1) node[above,yshift=3,scale=.8]{$0\ \mathrm{or}\ -$};
    \node[scale=.7,black,below right]at (p23) {$\alpha$};
    \node[scale=.7,black,xshift=4,yshift=4]at ($(p23)!.5!(p24)$) {$\alpha$};
    \node[scale=.7,black,xshift=5.5]at ($(p25)!.5!(p24)$) {$\alpha$};
\end{tikzpicture}
}}
\longrightarrow
\vcenter{\hbox{
\begin{tikzpicture}[scale=0.8,>=stealth]
    \def\p1{1}
    \coordinate (p1) at (0,1);
    \coordinate (p2) at (0.8660,-.5);
    \coordinate (p3) at (-0.8660,-.5);
    \node[below right] at (0,0) {$v$};
    \foreach \p in {1,2,3} {
        \foreach \i in {0,1,2,3,4,5} \coordinate (p\p\i) at ($(p\p)+(90-60*\i:1)$);
    }
    \draw[black,very thick](p23)--($(p23)+(0,-.3)$);
    \draw[black,very thick](p23)--(p24)--(p25);
    \draw[opacity=0.5,red,thick](p10)--(p11)--(p12)--(p13)--(p14)--(p15)--cycle;
    \draw[opacity=0.5,green,thick](p20)--(p21)--(p22)--(p23)--(p24)--(p25)--cycle;
    \draw[opacity=0.5,blue,thick](p30)--(p31)--(p32)--(p33)--(p34)--(p35)--cycle;
    \filldraw (0,0) circle (0.1);
    \filldraw (p10) circle (0.1);
    \filldraw (p20) circle (0.1);
    \filldraw (p23) circle (0.1);
    \filldraw (p30) circle (0.1);
    \filldraw (p33) circle (0.1);
    \filldraw (p32) circle (0.1);
    \node[scale=.7,red,left]at ($(p11)!.5!(p12)$) {$g_1$};
    \node[scale=.7,red,right]at ($(p14)!.5!(p15)$) {$-g_1$};
    \node[scale=.7,green,left]at ($(p21)!.5!(p22)$) {$g_2$};
    \node[scale=.7,blue,right]at ($(p34)!.5!(p35)$) {$-g_3$};
    \node[scale=.6,black,xshift=8]at ($(p12)!.5!(p13)$) {$g_1-g_2$};
    \node[scale=.6,black,xshift=-8]at ($(p13)!.5!(p14)$) {$g_3-g_1$};
    \node[scale=.6,black,xshift=8]at ($(p24)!.5!(p25)$) {$\alpha+g_3-g_2$};
    \node[scale=.6,black,xshift=5]at ($(p23)!.5!(p24)$) {$\alpha-g_2$};
    \node[scale=.7,blue,above left]at ($(p32)!.5!(p33)$) {$g_3$};
    \node[scale=.7,black,below right]at (p23) {$\alpha$};
\end{tikzpicture}
}}.
\end{align}
As the transformations on all the strings are known, we can use $q(a,b)$ in Eq.~(\ref{q}) to calculate the $\U$ charges of all other vertices except $v$. Using the global $\U$ charge conservation, the charge $Q_v$ of the right-hand-side configuration is
\begin{align}\nonumber
    Q_v&=(\delta_{[\alpha]}-1)
    +q(-g_1,g_1)
    +q(-g_3,g_3-g_1)
    +q(g_1-g_2,g_2)
    -q(-g_3,g_3)
    +q(g_3,\alpha-g_2)
    -q(\alpha-g_2,g_2)
    \\
    &=q_e(g_3-g_1,g_1-g_2,\alpha+g_3-g_2),
\end{align}
where $\delta_{[\alpha]}-1$ is the charge of vertex $v$ of the left-hand-side configuration (the charge is $0$ if $[\alpha]=0$, and $-1$ otherwise). And the six $q(a,b)$'s in the first line correspond to the charges of the six dotted vertices of the right-hand-side configuration except $v$. In the last line of the equation, we used the charge function
\begin{equation}\label{q_e}
    q_e(a,b,c)=\left\lfloor \frac{[a]+[b]-[c]}{n} \right\rfloor,
\end{equation}
which is introduced in Appendix~\ref{App:H}.
It is the decorated charge of the vertex $v$ with three edges labeled by $a,b$ and $c$ without the constraint $[c]=[a+b]$ in general. Compared to the charge $q$ in Eq.~(\ref{q}) of the ground state, $q_e$ has a floor function $\lfloor x\rfloor$, which is defined as the greatest integer less than or equal to $x$. The average value of the charge $Q_v$ is then
\begin{align}\nonumber
    \langle Q_v\rangle &= \frac{1}{n^3}\sum_{g_1,g_2,g_3\in \Z_n}
    q_e(g_3-g_1,g_1-g_2,\alpha+g_3-g_2)\\
    &=\delta_{[\alpha]}-1+\frac{n-1}{2n}+\dfrac{\alpha}{n}.
\end{align}
Therefore, the average $Q_v$ after the subtraction of the vacuum charge is
\begin{equation}\label{ZnQv}
    \langle Q_v\rangle - \langle Q_v\rangle_0 = \frac{\alpha}{n} \quad(\mathrm{mod}\ 1).
\end{equation}
It is exactly the expected fractional $\U$ charge of the endpoint of a string with group element label $\alpha$.

On the other hand, we can also define the cluster charge of our model by replacing $2$ in Eq.~\eqref{Qcluster} by $n$. In the same way as the $\Z_2$ toric code, we can show that the fractional charge of the fundamental $e$ anyon in $\Z_n$ toric code is $1/n$.

It will be useful to think of the model as a gauged $\U\times\Z_n$ SPT phase. The ``ungauged" model can be obtained from dualizing the $\Z_n$ strings: they become domain walls of $\Z_n$ spins living on the dual lattice, as illustrated below 
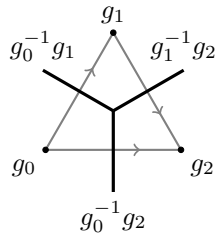
\begin{center}
\begin{tikzpicture}[scale=1.8]
\coordinate (0) at (0,0);
\coordinate (2) at (1,0);
\coordinate (1) at (.5,.866);
\coordinate (center) at (.5,.2886);
\draw[->-=.7,gray,thick](0)--(1);
\draw[->-=.7,gray,thick](1)--(2);
\draw[->-=.7,gray,thick](0)--(2);
\node[below left] at (0) {$g_0$};
\node[above] at (1) {$g_1$};
\node[below right] at (2) {$g_2$};
\coordinate (b) at ($(center)+(30:.6)$);
\coordinate (a) at ($(center)+(120+30:.6)$);
\coordinate (c) at ($(center)+(240+30:.6)$);
\draw[very thick](center)--(a);
\draw[very thick](center)--(b);
\draw[very thick](center)--(c);
\node[above] at (a) {$g_0^{-1}g_1$};
\node[above] at (b) {$g_1^{-1}g_2$};
\node[below] at (c) {$g_0^{-1}g_2$};
\filldraw (0) circle (0.02);
\filldraw (1) circle (0.02);
\filldraw (2) circle (0.02);
\node[] at (0,-.7) {};
\end{tikzpicture}
\end{center}
From this point of view, the vertex violations are actually $\Z_n$ symmetry fluxes. The fact that they carry fractional U(1) charge shows the underlying SPT phase is topologically nontrivial.

\section{Decorated string-net models}
We can turn towards the general case in 2D.
 In general, the topological order of a 2D gapped phase is described mathematically by a unitary modular tensor category (MTC) $\mathcal{B}$, which is an algebraic theory of anyon excitations at low energy. We denote the anyon types by $a, b, c,\dots$. The universal topological properties are completely captured by the fusion and braiding structures of the MTC. In particular, the fusion rules can be written as $a\times b=\sum_c N_{ab}^c c$, where integers $N_{ab}^c\geq 0$ are the fusion coefficients. 
 
\subsection{Review of string-net models}
All 2D topological phases with gappable boundaries have CPH realizations~\cite{freed2020gapped, Kitaev_2012}, known as the generalized string-net models~\cite{Levin_2005, lin2021generalized}.  We first briefly review the construction.

The model can be defined on any trivalent lattice, with spins on the edges. As discussed in Sec.~\ref{sec:general}, the lattice should be endowed with a branching structure so all edges are oriented. In all the diagrams below it is understood that each edge has an arrow pointing upwards. For each spin, there is an orthonormal basis, labeled by a finite set of ``string types". On each vertex, the configuration with the three edges with string types $a,b$ and $c$ is allowed if $a,b$ and $c$ satisfy the branching rule $N_{ab}^c=1$. Formally, the branching rules define associative fusion for the set of string types, similar to the fusion rules of anyons. With the vertex constraints imposed, there is still a large manifold of allowed string-net states. The ground state wavefunction is then a superposition of these states, and the amplitudes are defined by the following local relations: 
\begin{enumerate}
\item The wave function is invariant under local deformation of strings,
  \begin{equation}
    \label{eq:locrel1}
     \Psi\left(\begin{tikzpicture}[baseline={($ (current bounding box) - (0,0pt) $)}]
    \draw (0, .6)  --(0, -.6) [very thick] ;
    \draw (0, .3) ..controls(.3, .2)..(.3, .1)
    --(.3, -.1)
    ..controls(.3, -.2).. (0, -.3) [dashed, thick];
    \end{tikzpicture}\right)=
     \Psi\left(\begin{tikzpicture}[baseline={($ (current bounding box) - (0,0pt) $)}]
    \draw (0, .6)  --(0, .3) [very thick] ;
    \draw (0, .3) ..controls(.3, .2)..(.3, .1)
    --(.3, -.1)
    ..controls(.3, -.2).. (0, -.3) [ very thick];
    \draw (0, -.3)--(0, -.6) [very thick];
    \end{tikzpicture}\right).
  \end{equation}
Here the graph in the parenthesis represents a local patch of the string state. Here the dashed line represents the unique ``vacuum" string $0$, which satisfies $N_{0a}^a=N_{a0}^a=1$.
  
 \item  The wave function is invariant up to a normalization factor, under the creation/annihilation of bubbles:
  \begin{equation}
    \label{eq:locrel2}
    \Psi\left(\begin{tikzpicture}[baseline={($ (current bounding box) - (0,0pt) $)}]
      \draw (0, .6)--(0, .3) [very thick] node [midway, right] {$a$};
      \draw (0, .3) ..controls(-.3, .2)..(-.3, .1) --(-.3, -.1) node [midway,left] {$b$}
      ..controls(-.3, -.2).. (0, -.3) [very thick];
      \draw (0, .3) ..controls(.3, .2)..(.3, .1)
      --(.3, -.1) node [midway,right] {$c$}
      ..controls(.3, -.2).. (0, -.3) [very thick];
      \draw (0, -.3)--(0, -.6) [very thick]
	  node [midway, right] {$a^\prime$};
    \end{tikzpicture}\right)
    =\delta_{aa^\prime}\sqrt{\frac{d_{b}d_{c}}{d_{a}}}
    \Psi\left(\begin{tikzpicture}[baseline={($ (current bounding box) - (0,0pt) $)}]
    \draw (0, .6)--(0, .3) [very thick]
	node [midway, right] {$a$};
    \draw (0, .3) ..controls(-.3, .2)..(-.3, .1)
    --(-.3, -.1)
    ..controls(-.3, -.2).. (0, -.3) [very thick];
    \draw (0, .3) ..controls(.3, .2)..(.3, .1)
    --(.3, -.1)
    ..controls(.3, -.2).. (0, -.3) [dashed, thick];
    \draw (0, -.3)--(0, -.6) [very thick];
    \end{tikzpicture}\right)
    =\delta_{aa^\prime}\sqrt{\frac{d_{b}d_{c}}{d_{a}}}
    \Psi\left(\begin{tikzpicture}[baseline={($ (current bounding box) - (0,0pt) $)}]
    \draw (0, .6)--(0, .3) [very thick]
	node [midway, right] {$a$};
    \draw (0, .3) ..controls(-.3, .2)..(-.3, .1)
    --(-.3, -.1)
    ..controls(-.3, -.2).. (0, -.3) [dashed, thick];
    \draw (0, .3) ..controls(.3, .2)..(.3, .1)
    --(.3, -.1)
    ..controls(.3, -.2).. (0, -.3) [very thick];
    \draw (0, -.3)--(0, -.6) [very thick];
    \end{tikzpicture}\right).
  \end{equation}

\item The amplitudes satisfy the following F move when a local patch is reconnected:
	\begin{equation}
    \label{eq:fmove}
    \Psi\left(\begin{tikzpicture}[baseline={($ (current bounding box) - (0,0pt) $)}, scale=0.8]
		\draw [ very thick](0, 0) node [above] {$a$} --(.3, -.5) ;
		\draw [ very thick](.6, 0) node [above] {$b$} --(.3, -.5) ;
		\draw [ very thick](.3, -.5)--(.6, -1)  node [midway, left] {$e$}; 
		\draw [ very thick](1.2, 0) node [above] {$c$} --(.6, -1) ;
		\draw [ very thick](.6, -1)--(.9, -1.5)  node [below] {$d$};
		\end{tikzpicture}\right)
		=\sum_f [F^{abc}_d]_{ef}\:
    \Psi\left(\begin{tikzpicture}[baseline={($ (current bounding box) - (0,0pt) $)}, scale=0.8]
		\draw [ very thick](0, 0) node [above] {$a$} --(.6, -1) ;
		\draw [ very thick](.6, 0) node [above] {$b$} --(.9, -.5) ;
		\draw [ very thick](.9, -.5)--(.6, -1) 
		      node [midway, right] {$f$};
		\draw [ very thick](1.2, 0) node [above] {$c$} --(.9, -.5) ;
		\draw [ very thick](.6, -1)--(.9, -1.5)  node [below] {$d$};
		\end{tikzpicture}\right).
  \end{equation}
  The $F$ symbols must satisfy the pentagon identity in order to have consistent local moves.
\end{enumerate}
Together these datum form a unitary fusion category (UFC), which will be denoted by $\mathcal{C}$ in the following. 

The wavefunction is the ground state of the following CPH:
\begin{equation}
    H=-\sum_v A_v-\sum_p B_p,
\end{equation}
where $A_v$ is defined on each vertex $v$ essentially imposing the branching rules energetically, and the plaquette term $B_p$ introduces fluctuations to the string-net states locally and selects the desired ground state. For more details about the Hamiltonian, we refer the readers to Ref. [\onlinecite{lin2021generalized}]. The resulting topological order, i.e. the MTC, is mathematically the Drinfeld center $\mathcal{Z}(\mathcal{C})$ of the UFC $\mathcal{C}$. It can be shown that a topological phase admits a string-net realization (i.e. the MTC is a Drinfeld center) if and only if the topological phase can have a fully gapped boundary.

For reasons that will become clear shortly, we assume that the input UFC is graded by a group $G$.
  Namely, the labels can be divided into subsets $\mathcal{C}_\mb{g}$ for $\mb{g}\in G$, such that
\begin{equation}
    a_\mb{g}\times b_\mb{h}=\sum_{c_\mb{gh}\in \mathcal{C}_\mb{gh}}c_\mb{gh}.
\end{equation}
Such a string-net model can be understood as a topological $G$ gauge theory: it can be dualized to a string-net model with input $\mathcal{C}_1$ (the identity component), enriched by global symmetry $G$. 
 In particular, vertex excitations carry symmetry flux of the $G$ gauge field, and the flux is just the violation of the $G$-grading at the vertex. For instance, the following vertex
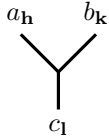
\begin{center}
\begin{tikzpicture}[scale=0.25,>=stealth]
    \def\shift{0};
    \def\shifty{0}
	\draw[very thick] (\shift+0,\shifty+4) node[above]{$a_\mb{h}$}  -- (\shift+2,\shifty+2) ;
	\draw[ very thick]  (\shift+4,\shifty+4) node[above]{$b_\mb{k}$} -- (\shift+2,\shifty+2) ;
	\draw[ very thick](\shift+2,\shifty+2) -- (\shift+2,\shifty+0) node[below]{$c_\mb{l}$};
	\end{tikzpicture}
\end{center}
has a flux $\mb{g}=\mb{h}\mb{k}\mb{l}^{-1}$.

It was shown in Ref. [\onlinecite{Cheng_2017}] and [\onlinecite{Heinrich_2016}], based on the mathematical results in Ref. [\onlinecite{ENO}] that any anomaly-free $G$ symmetry-enriched $\mathcal{Z}(\mathcal{C}_1)$ phase can be realized by a $G$-graded extension of $\mathcal{C}_1$.

\subsection{Charge decoration}
Now we define a string-net wavefunction with U(1) charge decoration, assuming the input is a $G$-graded UFC. Consider a group homomorphism from $G$ to $\U$, and write the image of $\mb{g}\in G$ as $e^{2\pi i m(\mb{g})}$, where $m(\mb{g})\in[0,1)$. Define
\begin{equation}
    q_{\mb{g,h}}=m(\mb{g})+m(\mb{h})-m({\mb{gh}}).
\end{equation}
Clearly $q_\mb{g,h}\in\Z$. Since $0\leq m(\mb{g})<1$, we have $-1< q_\mb{g,h}<2$, which means $q_\mb{g,h}=0,1$.  Thus we place at each vertex a spin-$1$ with U(1)$_{S_z}$ symmetry. We modify the vertex constraint to impose the following decoration rule:
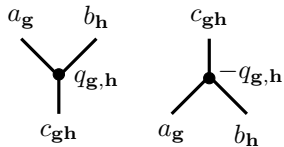
\begin{center}
\begin{tikzpicture}[scale=0.25,>=stealth]
    \def\shift{0};
    \def\shifty{0}
	\draw[very thick] (\shift+0,\shifty+4) node[above]{$a_\mb{g}$}  -- (\shift+2,\shifty+2) ;
	\draw[ very thick]  (\shift+4,\shifty+4) node[above]{$b_\mb{h}$} -- (\shift+2,\shifty+2) ;
	\draw[ very thick](\shift+2,\shifty+2) -- (\shift+2,\shifty+0) node[below]{$c_\mb{gh}$};
    \filldraw (\shift+2, \shifty+2.1) circle (0.3) ;
	\node at (\shift+4, \shifty+1.8) {$q_{\mb{g,h}}$};
    
    \def\shift{8};
    \def\shifty{0}
	\draw[ very thick] (\shift+0,\shifty+0) node[below]{$a_\mb{g}$}  -- (\shift+2,\shifty+2) ;
	\draw[ very thick]  (\shift+4,\shifty+0) node[below]{$b_\mb{h}$} -- (\shift+2,\shifty+2) ;
	\draw[ very thick](\shift+2,\shifty+2) -- (\shift+2,\shifty+4) node[above]{$c_\mb{gh}$};
    \filldraw (\shift+2, \shifty+1.9) circle (0.3);
    \node at (\shift+4.2, \shifty+2.1) {$-q_\mb{g,h}$};
    \end{tikzpicture}

\end{center}
We can easily check that U(1) charge is preserved by all the local relations. Together we have defined a string-net wavefunction with U(1)$_{S_z}$ symmetry, and it is not difficult to modify the Hamiltonian to take care of the charge decoration, similar to what has been done in the $\Z_n$ case in Sec. \ref{sec:Zn}. As we will see later, it is sufficient to consider $G$ being a finite cyclic group, in which case the vertex degree of freedom is a spin-1.

The $\Z_n$ toric code defined in Sec. \ref{sec:Zn} is an example of the decorated string-net construction. The label set is basically the $\Z_n$ group (denoted additively): $\{[0],[1], \cdots, [n-1]\}$, with the branching rule given by the mod $n$ addition. The $F$ symbol is set to $1$ for any permissible move. The $q$ function is defined from $m(a)=\frac{[a]}{n}$ for $a\in \Z_n$.

 A useful way to think about the model is the following coupled bilayer construction: without loss of generality, take $G=\Z_N$. Considers two layers of aligned honeycomb lattice models. One layer is the string-net model with the input being the $\Z_N$-graded UFC $\mathcal{C}$, and the other layer is the model presented in Sec. \ref{sec:Zn}. We then turn on the coupling between the two layers, locking their $\Z_N$ gradings. Assuming that the locking is the highest energy scale, compared with the interaction strengths of the terms in the string-net models, the coupling essentially decorates the charge to the string-net model, yielding exactly the model introduced in this section. Physically, one can think of the inter-layer coupling as a Higgs interaction that identifies the two $\Z_N$ gauge fields. Furthermore, if both layers are ``ungauged", or in the lattice model the gauge fields dualized to spins, the bilayer system can be regarded as a $\U\times\Z_N$ SPT phase, stacked on a $\Z_N$-enriched string-net model (without any charge decoration). After gauging, anyons carrying flux $\mb{g}$ now has fractional charge given exactly by $m(\mb{g})$. 

Below we show that the construction presented in this section includes the most general U(1) symmetry-enriched topological phases that admit possible CPH realizations. 

First we review the general classification of U(1) symmetry-enriched topological phases~\cite{SET}.
We will assume that local excitations carry integer charges $0,1,\dots$. While all anyon excitations are charged, modulo local charged excitations. We can associate to each anyon type $a$ a fractional charge $q_a\in[0,1)$, which is a universal topological quantum number. Importantly, the fractional charge assignment should be compatible with fusion rules:
\begin{equation}
    e^{2\pi iq_a} e^{2\pi i q_b} = e^{2\pi i q_c}, \text{ if }N_{ab}^c>0.
\end{equation}
Then as shown in Ref. [\onlinecite{SET}] there exists an Abelian anyon $\coho{v}$ such that $e^{2\pi i q_a}=M_{\cohosub{v}a}^*$, where $M_{\cohosub{v}a}$ is the braiding phase between $\coho{v}$ and $a$. In the following $\coho{v}$ will be referred to as the vison. Physically, $\coho{v}$ is the anyon created by adiabatically inserting $2\pi$ U(1) flux. The braiding between $\coho{v}$ and an anyon of type $a$ is thus equal to the Aharonov-Bohm phase $e^{i 2\pi q_a}$. By Laughlin's argument, the charge carried by $\coho{v}$ itself is essentially the Hall conductance measured in units of $\frac{e^2}{h}$. In fact, we have a stronger result $e^{i\pi \sigma_H}=\theta_{\cohosub{v}}$, proven in Ref. [\onlinecite{Kapustin_2020}]. The physical argument given here reproduces the mathematical classification in Ref. [\onlinecite{SET}], that the symmetry fractionalization class for a group $G$ is given by $\H^2[G, \mathcal{A}]$, where $\mathcal{A}$ is the fusion group of Abelian anyons. In this case, $\H^2[\U, \mathcal{A}]=\mathcal{A}$.

As shown in Ref. [\onlinecite{Kapustin_2019}], a commuting projector Hamiltonian must have zero Hall conductance. Together with $e^{i\pi\sigma_H}=\theta_{\cohosub{v}}$, it implies that the vison $\coho{v}$ must be a boson. A bosonic vison strongly constrains the structure of the topological order: The theory $\mathcal{B}$ can be reduced by condensing the boson $\coho{v}$, which confines all anyons braiding nontrivially with $\coho{v}$, or equivalently those with fractional charges. This process does not break charge conservation since $\coho{v}$ is charge-neutral, and the remaining topological order $\mathcal{B}_0$ is also charge-neutral, i.e. all anyons carry only integer charges. It is known that the MTC $\mathcal{B}$ and $\mathcal{B}_1$ are related by gauging: if the order of $\coho{v}$ is $N$, then $\mathcal{B}$ can be obtained by gauging a $\Z_N$ symmetry in $\mathcal{B}_1$. It then follows that the $\Z_N$ symmetry fluxes are fractionally charged. 

Since by our assumption $\mathcal{B}$ can be realized by a string-net model, mathematically $\mathcal{B}$ must be equivalent to the Drinfeld center of a UFC $\mathcal{C}$, which is the input to the string-net construction. Furthermore, as shown in Ref. [\onlinecite{Cheng_2017}], if $\mathcal{B}$ can be obtained from $\mathcal{B}_1$ by gauging a $\Z_N$ symmetry, $\mathcal{C}$ must be a $\Z_N$-graded fusion category. In fact, $\mathcal{B}_1$ must be the Drinfeld center of the identity component $\mathcal{C}_1$.

Altogether, we have shown that the construction presented in this section is capable of realizing any 2D U(1) symmetry-enriched topological phase, as long as the topological order is a Drinfeld center and the Hall conductance vanishes.

\section{Three-dimensional generalizations}
We now consider how our construction can be generalized to three dimensions to realize topologically ordered phases with U(1) symmetry fractionalization.

\subsection{Decorated Walker-Wang models}
\begin{figure}[h]
\centering
\subfloat[]{\label{fig:WWa}
\begin{tikzpicture}[scale=3]
\coordinate (2) at (0,0);
\coordinate (0) at (.2,-.4);
\coordinate (1) at (1,0);
\coordinate (3) at (.5,.7);
\draw[->-=.8,gray,dashed,thick](1)--(2);
\draw[->-=.6,gray,thick](2)--(3);
\coordinate (012) at (.4,-.5);
\coordinate (023) at (0,.45);
\coordinate (02) at (.4,-.05);
\coordinate (023-x) at ($(023)!.58!(02)$);
\draw[->-=.6,very thick,black](023)--(023-x);
\draw[very thick,densely dotted,black](023-x)--(02);
\coordinate (012-x1) at ($(012)!.36!(02)$);
\coordinate (012-x2) at ($(012)!.75!(02)$);
\draw[->-=.8,very thick,black](012)--(012-x1);
\draw[very thick,densely dashed,black](012-x1)--(012-x2);
\draw[very thick,densely dotted,black](012-x2)--(02);
\coordinate (013) at (1,.45);
\coordinate (123) at (0.65,.8);
\coordinate (13) at (.55,.15);
\coordinate (123-x1) at ($(13)!.24!(123)$);
\coordinate (123-x2) at ($(13)!.65!(123)$);
\draw[very thick,densely dotted,black](13)--(123-x1);
\draw[very thick,densely dashed,black](123-x1)--(123-x2);
\draw[->-=.5,very thick,black](123-x2)--(123);
\draw[->-=.6,gray,thick](1)--(3);
\coordinate (013-x) at ($(013)!.75!(13)$);
\draw[very thick,densely dotted,black](13)--(013-x);
\draw[->-=.65,very thick,black](013-x)--(013);
\draw[->-=.6,very thick,densely dotted,black](02)--(13);
\draw[->-=.6,gray,thick](0)node[scale=.8,below,black]{0}--(1)node[scale=.8,right,black]{1};
\draw[->-=.6,gray,thick](0)--(2)node[scale=.8,left,black]{2};
\draw[->-=.6,gray,thick](0)--(3)node[scale=.8,above,black]{3};
\fill [black] (02) circle (.7pt);
\fill [black] (13) circle (.7pt);
\node[below] at (012) {$x_{012}$};
\node[right] at (013) {$x_{013}$};
\node[above] at (023) {$x_{023}$};
\node[above] at (123) {$x_{123}$};
\node[right] at ($(02)!.5!(13)$) {$x_{0123}$};
\node[xshift=-6,yshift=3] at (13) {$v_1$};
\node[xshift=7,yshift=-3] at (02) {$v_2$};
\end{tikzpicture}
}
\subfloat[]{\label{fig:WWb}
\begin{tikzpicture}[scale=.9]
\coordinate (a) at (-1,1.7);
\coordinate (c) at (1,1.7);
\coordinate (b) at (-1,-2);
\coordinate (d) at (1,-2);
\coordinate (x1) at (0,.7);
\coordinate (x2) at (0,-1);
\draw[->-=.6,very thick](b)--(x2);
\draw[->-=.6,very thick](d)--(x2);
\draw[->-=.6,very thick](x2)--(x1);
\draw[->-=.6,very thick](x1)--(a);
\draw[->-=.6,very thick](x1)--(c);
\fill [black] (x1) circle (2.3pt);
\fill [black] (x2) circle (2.3pt);
\node[right] at (x1) {$q(a,c)$};
\node[right] at (x2) {$-q(b,d)$};
\node[below left] at (x1) {$v_1$};
\node[above left] at (x2) {$v_2$};
\node[above] at (a) {$a=x_{123}$};
\node[below] at (b) {$b=x_{023}$};
\node[above] at (c) {$c=x_{013}$};
\node[below] at (d) {$d=x_{012}$};
\node[right,xshift=2] at ($(x1)!.5!(x2)$) {$e=x_{0123}$};
\end{tikzpicture}
}
\caption{$\U$ charge decorations for Walker-Wang model. (a) The four triangle faces of a tetrahedron are dual to the strings $x_{ijk}\in\mathcal{B}$. The string joint point inside the tetrahedron is further resolved into two trivalent vertices $v_1$ and $v_2$. (b) The dual string trivalent lattice. The vertices $v_1$ and $v_2$ are decorated by $\U$ charge $q(a,c)$ and $-q(b,d)$, respectively. So the total charge inside the tetrahedron is $q_{3D}(a,b,c,d)=q(a,c)-q(b,d)$ as Eq.~(\ref{q3D}).}
\end{figure}
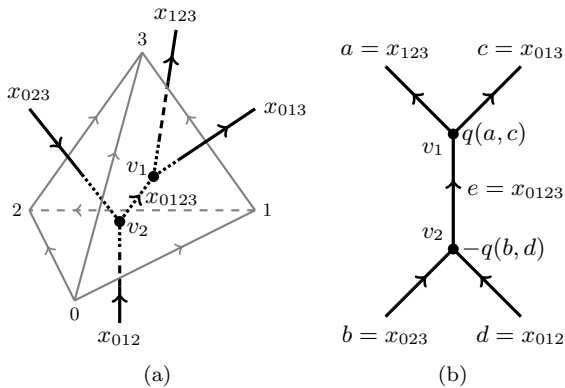

 We will start from a class of three-dimensional CPH models, known as the Crane-Yetter-Walker-Wang (CYWW) models~\cite{walker201131tqfts}, which can be thought of as a natural generalization of the string-net models to 3D. The input to the CYWW construction is a braided fusion category (BFC), roughly speaking an anyon model but the modularity condition does not have to be imposed. The model can be defined on any triangulation, where each triangle face and tetrahedron is attached a Hilbert space with basis states labeled by anyon types in the input BFC. It is very useful to think about the dual triangulation, where the states are now on dual edges (``strings"), and the allowed configurations are defined by the fusion rules at each vertex of the dual triangulation (see Fig.~\ref{fig:WWa}).
The ground state wavefunction is again a superposition of all permissible string configurations. The amplitude for a configuration is given by the evaluation of the anyon diagram obtained from projecting the 3D configuration to a certain plane. We refer the readers to Refs. [\onlinecite{walker201131tqfts}, \onlinecite{von_Keyserlingk_2013}] for more details about the CPH. 

As shown in Ref. [\onlinecite{walker201131tqfts}], when the input BFC is modular, the ground state is short-range entangled. When there is a surface, one finds a 2D topological phase described by exactly the input MTC. In a way, the 3D Walker-Wang model provides a CPH realization of any 2D MTC, including chiral ones, at the expense of introducing a 3D trivial bulk. When the input BFC is not modular, one can show that there must be a subset of anyon labels closed under fusion, i.e. a subcategory, which has trivial braiding with every other anyon. This subcategory is known as the M\"uger center. A theorem due to Degline shows that the M\"uger center is always isomorphic to the representation category Rep($H$) for some finite group $H$~\footnote{More precisely, Rep($H,z$) where $z$ is a $\Z_2$ involution on $H$. The effect of $z$ is to make some of the particles in the representation category fermionic}. In this case, the Walker-Wang model describes a $H$ gauge theory~\cite{von_Keyserlingk_2013}, possibly twisted~\cite{Wang_2017}. 

Now we turn to symmetry enrichment. Symmetry-enriched generalizations of the CYWW modelas have been proposed in Ref. [\onlinecite{BulmashPRR2020}] for finite group symmetry, see also [\onlinecite{Williamson_2017}]. Here we will focus on U(1) charge decoration. Similar to the 2D case, we assume that the input BFC is graded by a finite group $G$, and choose a homomorphism from $G$ to $\mathbb{R}/\Z$. For simplicity we will just write $m(x)$ in the following for $x\in \mathcal{B}$. One way to construct such a map is to choose an Abelian anyon $a\in \mathcal{B}$, and for $x\in \mathcal{B}$ define $e^{2\pi i m(x)}=M_{ax}^*$. Clearly, in this case $a$ can be interpreted as the vison, and if $\mathcal{B}$ is modular these are all the possibilities. Define \begin{equation}
    q(x,y)=m(x)+m(y)-m(x\times y).
\end{equation}
Here $x\times y$ means any anyon in the fusion outcome of $x$ and $y$ as they should all have the same value of $m$ by the grading.

In the CYWW model, we introduce an additional spin in the center of the tetrahedron. Suppose the four vertices are ordered as $0123$.  The string dual to a triangle face $\langle ijk\rangle$ of a triangulation is labeled by $x_{ijk}\in\mathcal{B}$, with orientation given by the right-hand rule (see Fig.~\ref{fig:WWa}).
Every joint point of the four strings (i.e., the tetrahedron center) is resolved into two vertices inside the tetrahedron. The dual graph is then a trivalent lattice. To clarify the geometric positions, the dual lattice for a tetrahedron is redrawn in Fig.~\ref{fig:WWb}. It is very similar to a unit cell of a 2D honeycomb lattice discussed in previous sections. We will use the abbreviations $a=x_{123},b=x_{023},c=x_{013},d=x_{012}$ and $e=x_{0123}$. 

Inspired by the 2D construction, we put charges $q(a,c)$  and $-q(b,d)$ to the two vertices inside the tetrahedron. Then the total charge of the tetrahedron is
\begin{equation}\label{q3D}
    q_{\mathrm{3D}}(a,b,c,d)=q(a,c)-q(b,d)=m(a)-m(b)+m(c)-m(d).
\end{equation}
Alternatively, if one directly works with the dual triangulation, charges can be decorated to vertices following the above rule.


Let us consider a simple example of the charge fractionalization in the Walker-Wang model with the symmetric category $\mathcal B=\mathrm{Vec}_{\Z_n}$. The string types are labeled by $x=0, 1,\dots, n-1$ and we denote $x\text{ mod }n$ by $[x]$. Since the M\"uger center of $\mathrm{Vec}_{\Z_n}$ is itself, this CYWW model describes a 3D (untwisted) $\Z_n$ gauge theory. 
In the ground state, the string configuration of every trivalent vertex should satisfy the fusion rule of $\Z_n$. If we choose $m(x)=\frac{[x]}{n}$, then the total charge inside a tetrahedron is
\begin{equation}\label{q3DZn}
    q_\mathrm{3D}(a,b,c,d)=\frac{[a]-[b]+[c]-[d]}{n},
\end{equation}
where we have the fusion constraint $[a-b+c-d]=0$. In fact, the above $q_\mathrm{3D}$ is the generating 3-cocycle in $\H^3[B^2\Z_n,\Z]=\Z_n$. 
Since an outgoing (ingoing) string contributes $[x]/n$ ($-[x]/n$) to the $\U$ charge of a tetrahedron, the contributions of a single string for its two neighboring tetrahedra cancel each other. Therefore, the charges of all the tetrahedra in an arbitrary triangulation of a closed 3D spacial manifold will sum up to zero. As a result, our decoration rule preserves the global $\U$ charge symmetry.

The fractional charge of the point-like $e$ excitation in the CYWW model can be calculated similarly to the 2D case. The plaquette operator $B_{\langle ij \rangle}^{(g_{ij})}$ dual to the link $\langle ij\rangle$ changes the labels of the strings around this link by $g_{ij}\in\Z_n$ (using the right-hand rule). For a tetrahedron shown in Fig.~\ref{fig:WWa}, there are in total $6$ plaquette operators dual to the $6$ links of the tetrahedron. After the action of these operators, the string labels of Fig.~\ref{fig:WWb} become
\begin{equation}
\begin{aligned}
    a\ &\rightarrow\  a'=a+g_{12}-g_{13}+g_{23},\\
    b\ &\rightarrow\  b'=b+g_{02}-g_{03}+g_{23},\\
    c\ &\rightarrow\  c'=c+g_{01}-g_{03}+g_{13},\\
    d\ &\rightarrow\  d'=d+g_{01}-g_{02}+g_{12},\\
    e\ &\rightarrow\  e'=e+g_{01}-g_{03}+g_{12}+g_{23}.
\end{aligned}
\end{equation}
The ground state is an equal-weight superposition of all these allowed configurations. So the average vacuum charges of the two vertices inside the tetrahedron can be calculated by averaging the $g_{ij}$'s as
\begin{equation}
    \langle Q_{v_1}\rangle_0 = -\langle Q_{v_2}\rangle_0 = \frac{1}{n^6} \sum_{\{g_{ij}\in\Z_n\}} q(a',c') = \frac{n-1}{2n}.
\end{equation}
Since the action of the plaquette operators around a vertex is similar to the 2D case, the above results are also the same as Eq.~\eqref{Qv0Zn}. The total vacuum charge of the tetrahedron $t$ is $\langle Q_{t}\rangle_0=\langle Q_{v_1}\rangle_0+\langle Q_{v_2}\rangle_0=0$.

The point-like $e$ excitations of the model are the gauge charges violating the $\Z_n$ fusion rule at some vertices. For string configurations with $e$ excitations, we have to change the $\U$ charge decoration rule to $q_e(a,b,c)$ defined in Eq.~(\ref{q_e}). The $\U$ charge decoration of a tetrahedron is then
\begin{equation}
    q_{\mathrm{3D}}^e(a,b,c,d,e)
    =q_e(a,c,e)-q_e(b,d,e)
    =\left\lfloor \frac{[a]+[c]-[e]}{n} \right\rfloor
    -\left\lfloor \frac{[b]+[d]-[e]}{n} \right\rfloor,
\end{equation}
for arbitrary $a,b,c,d,e\in\Z_n$. We note that $q_{\mathrm{3D}}^e(a,b,c,d,e)$ reduces to $q_{\mathrm{3D}}(a,b,c,d)$ of Eq.~(\ref{q3DZn}) in the $e$-charge-free subspace with string labels $[e]=[a+c]=[b+d]$. If there is an $e$ excitation labeled by $\alpha\in\Z_n$ located at $v_1$, we can calculate the average charge of the tetrahedron as
\begin{equation}
    \langle Q_t\rangle = \frac{1}{n^6} \sum_{\{g_{ij}\}\in\Z_n} q_\mathrm{3D}^e(a',b',c',d',e') = \frac{\alpha}{n}
    \quad(\mathrm{mod}\ 1),
\end{equation}
in the subspace with $[e-a-c]=\alpha$ and $[b+d-e]=0$. We conclude that the $e$ excitation has fractional $\U$ charge similar to the 2D case.

It is interesting to consider the case of a modular input category $\mathcal{B}$. Although the bulk is completely trivial, the boundary is a 2D topological order described by $\mathcal{B}$. Now with the U(1) decorations, it is natural to conjecture that surface anyons carry fractional charges exactly given by the vison $\coho{v}$. Crucially, now $\coho{v}$ does not need to be bosonic. So in a sense at the price of a 3D trivial bulk, any 2D U(1)-enriched topological order, including those with non-zero Hall conductance, can be realized on the surface of a 3D decorated Walker-Wang model. 

As a particularly interesting application, if we choose $\mathcal{B}$ to be the $\Z_2$ topological order, and set $\coho{v}=\psi$, the surface realizes the so-called eCmC topological order~\cite{Vishwanath2013}. Since all the topological data of the $\Z_2$ toric code are real, the bulk Hamiltonian is invariant under complex conjugation, i.e. time-reversal invariant ($S_z$ does not transform under time reversal). We conclude that the bulk is a bosonic topological insulator protected by $\U\rtimes\Z_2^\TT$ symmetry, with $\Theta=2\pi$~\cite{Metlitski_2013}. 
Interestingly, the bosonic TI phase can also be understood as a superposition of time-reversal domain walls decorated by 2D bosonic quantum Hall states. Since 2D bosonic quantum Hall states can not have CPH, this domain wall picture can not be realized in exactly solvable model. Here instead we use decorated WW models to circumvent the problem. Similar ideas have been considered in Ref. [\onlinecite{Chen_2014}] for electronic TI phases.

\subsection{Decorated Dijkgraaf-Witten models}
In general, a (3+1)d topological phase has both particle and loop excitations. It is also possible for loop excitations to have ``fractionalization" of the global symmetry. The full classification of symmetry fractionalization on loop excitations is not currently known. Below we consider loop symmetry fractionalization in a twisted $G$ gauge theory, assuming that the gauge charges are all bosonic. If the gauge charges do not have fractional U(1) charges, then all such topological phases can be obtained from partially gauging a $\U\times G$ bosonic SPT phase. Applying K\"nneth formula, one finds two ``cross" terms
\begin{equation}
\begin{split}
  \H^3[G, \H^1[\U, \U]]&=\H^3[G, \Z]\simeq \H^2[G, \U]\\
  \H^1[G, \H^3[\U, \U]]&=\H^1[G, \Z].
\end{split}
\end{equation}
For a finite $G$, $\H^1[G, \Z]$ is trivial. We are therefore left with $\H^3[G, \Z]$ to describe nontrivial U(1) fractionalization on loop excitations once $G$ is gauged. For an example, consider $G=\Z_n\times\Z_n$. Denote the elements of $G$ by $\mb{a}=(a_1,a_2)$ where $a_1,a_2\in \{0,1,\dots,n-1\}$. The multiplication in $G$ is denoted additively: $(a_1,a_2)\cdot (b_1,b_2)=([a_1+b_1],[a_2+b_2])$. The generating $3$-cocycle in $\H^3[G, \Z]$ is given by
\begin{equation}\label{qabc}
    q_{\mathrm{3D}}(\mb{a},\mb{b},\mb{c})=\frac{1}{n}([a_1b_2]+[(a_1+b_1)c_2]-[a_1(b_2+c_2)]-[b_1c_2]).
\end{equation}
Mathematically, this is the Bockstein of the $2$-cocycle in $\H^2[G, \U]$. The K\"unneth decomposition $\H^3[G, \H^1[\U,\U]]$ suggests that the theory can be obtained from U(1) charge decoration on 0D junctions of $G$ symmetry fluxes. We will present such a construction below.

   First, we start from the lattice model of a twisted $G$ gauge theory on a triangulation~\cite{DW, Mesaros_2013, JiangPRX2014}. $G$ gauge fields live on edges of the lattice, and we impose the condition that there is no flux on any face. Each tetrahedron then has three independent gauge fields, $\mb{g}_{01}, \mb{g}_{12}, \mb{g}_{23}$. The ground state wavefunction is a superposition of all allowed configurations, possibly weighted with a phase factor determined by a $\H^4[G, \U]$ cocycle. 

Therefore, given a 3-cocycle $q_{\mathrm{3D}}(\mb{g,h,k})$, we now place a spin at the center of the tetrahedron. Fixing the gauge field configuration of the tetrahedron, the center spin is in the state with $S^z=q_{\mathrm{3D}}(\mb{g}_{01}, \mb{g}_{12}, \mb{g}_{23})$. The U(1) symmetry is again generated by the total $S^z$.

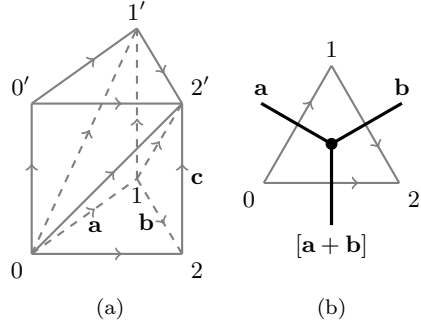
\begin{figure}[h]
\centering
\subfloat[]{\label{fig:slant1}
\begin{tikzpicture}[scale=2]
\coordinate (0) at (0,0);
\coordinate (2) at (1,0);
\coordinate (1) at (.7,.5);
\coordinate (0b) at (0,1);
\coordinate (2b) at (1,1);
\coordinate (1b) at (.7,1.5);
\draw[->-=.6,gray,dashed,thick](0)--(1);
\draw[->-=.6,gray,dashed,thick](1)--(2);
\draw[->-=.6,gray,thick](0)--(2);
\draw[->-=.6,gray,thick](0b)--(1b);
\draw[->-=.6,gray,thick](1b)--(2b);
\draw[->-=.6,gray,thick](0b)--(2b);
\draw[->-=.6,gray,thick](0)--(0b);
\draw[->-=.4,gray,dashed,thick](1)--(1b);
\draw[->-=.6,gray,thick](2)--(2b);
\draw[->-=.55,gray,thick](0)--(2b);
\draw[->-=.55,gray,dashed,thick](0)--(1b);
\draw[->-=.5,gray,dashed,thick](1)--(2b);
\node[below left] at (0) {$0$};
\node[below] at (1) {$1$};
\node[below right] at (2) {$2$};
\node[above,xshift=-4] at (0b) {$0'$};
\node[above] at (1b) {$1'$};
\node[above,xshift=7] at (2b) {$2'$};
\node[xshift=4,yshift=-4] at ($(0)!.5!(1)$) {$\mb{a}$};
\node[xshift=-5,yshift=-2] at ($(1)!.5!(2)$) {$\mb{b}$};
\node[right] at ($(2)!.5!(2b)$) {$\mb{c}$};
\end{tikzpicture}
}
\subfloat[]{\label{fig:slant2}
\begin{tikzpicture}[scale=1.8]
\coordinate (0) at (0,0);
\coordinate (2) at (1,0);
\coordinate (1) at (.5,.866);
\coordinate (center) at (.5,.2886);
\draw[->-=.7,gray,thick](0)--(1);
\draw[->-=.7,gray,thick](1)--(2);
\draw[->-=.7,gray,thick](0)--(2);
\node[below left] at (0) {$0$};
\node[above] at (1) {$1$};
\node[below right] at (2) {$2$};
\coordinate (b) at ($(center)+(30:.6)$);
\coordinate (a) at ($(center)+(120+30:.6)$);
\coordinate (c) at ($(center)+(240+30:.6)$);
\draw[very thick](center)--(a);
\draw[very thick](center)--(b);
\draw[very thick](center)--(c);
\node[above] at (a) {$\mb{a}$};
\node[above] at (b) {$\mb{b}$};
\node[below] at (c) {$[\mb{a}+\mb{b}]$};
\filldraw (center) circle (0.04);
\node[] at (0,-.7) {};
\end{tikzpicture}
}
\caption{Dimension reduction and slant product. (a) The 3D prism is triangulated into three tetrahedra, decorated with charges $q_{\mathrm{3D}}(\mb{a},\mb{b},\mb{c})$, $-q_{\mathrm{3D}}(\mb{a},\mb{c},\mb{b})$ and $q_{\mathrm{3D}}(\mb{c},\mb{a},\mb{b})$, respectively. (b) After the dimension reduction, the decorated charge associated to the triangle $\langle 012\rangle$ is $(\iota_\mb{c} q_{\mathrm{3D}})(\mb{a},\mb{b})$.}
\end{figure}

In the following, we will illustrate the symmetry fractionalization of the loop excitations in an explicit example of $G=\Z_n\times\Z_n$. We will show that the linked flux loops of the two $\Z_n$ gauge groups host fractional $\U$ charge. 
Let us consider the dimension reduction of a 3D lattice with a small size along the periodic $z$ direction~\cite{WangPRL2014}. If we have a loop excitation of the gauge flux $\mb{c}\in G$ created by the membrane operator parallel to the 2D plane, all the edges of the triangulation along the $z$ direction (crossing the membrane operator) will have gauge fields $\mb{c}\in G$. The building blocks of the 3D triangulation are prisms shown in Fig.~\ref{fig:slant1}. Each prism can be triangulated into three tetrahedra, $0122', 011'2'$ and $00'1'2'$, decorated with $\U$ charges $q_{\mathrm{3D}}(\mb{a},\mb{b},\mb{c})$, $-q_{\mathrm{3D}}(\mb{a},\mb{c},\mb{b})$ and $q_{\mathrm{3D}}(\mb{c},\mb{a},\mb{b})$, respectively. The second charge has a minus sign, since the associated tetrahedron has a different orientation. 
After the dimension reduction, the prism becomes a triangle of the 2D lattice shown in Fig.~\ref{fig:slant2}. And the $\U$ charge decorated at the center of the 2D triangle is the total charge of the three tetrahedra in 3D. The result is called the slant product of the 3-cocycle $q_{\mathrm{3D}}$:
\begin{equation}
\begin{aligned}
    (\iota_\mb{c} q_{\mathrm{3D}})(\mb{a},\mb{b})
    &:= q_{\mathrm{3D}}(\mb{a},\mb{b},\mb{c})-q_{\mathrm{3D}}(\mb{a},\mb{c},\mb{b})+q_{\mathrm{3D}}(\mb{c},\mb{a},\mb{b})\\
    &=c_1\cdot q_{\mathrm{2D}}(a_2,b_2) -c_2\cdot q_{\mathrm{2D}}(a_1,b_1),
\end{aligned}
\end{equation}
where $q_{\mathrm{3D}}$ and $q_{\mathrm{2D}}$ are defined in Eqs.~(\ref{qabc}) and (\ref{q}), respectively. If we choose the base flux $\mb{c}$ to be the fundamental flux $(c_1,c_2)=(1,0)$ of the first $\Z_n$, then the 2D charge decorated in Fig.~\ref{fig:slant2} would be $q_{\mathrm{2D}}(a_2,b_2)$.
As discussed in Sec.~\ref{sec:Zn}, this is exactly the 2D $\U$ charge decorated $\Z_n$ topological order for the second $\Z_n$ of $G$. When the flat connection condition is violated in Fig.~\ref{fig:slant2} by another flux perpendicular to the 2D plane, the center of the triangle will host a fractional $\U$ charge according to Eq.~(\ref{ZnQv}). Therefore, the mutually linked flux loops of the two $\Z_n$ gauge groups have fractional $\U$ charges.

\section{Discussions}
In this work, we present a general strategy to construct exactly solvable models for $\U$ symmetry enriched topological phases in both 2D and 3D. In 2D, our results show that as long as there are no protected gapless edge modes (either by the topological order or the U(1) symmetry, i.e. nonzero Hall conductance), the phase can be realized by a CPH. In 3D, we present examples of CPHs where U(1) symmetry fractionalize on particles or loops. The former is realized through a decorated Walker-Wang model, which also provides a CPH model for bosonic topological insulator. 

Compared to existing constructions, our model requires only finite-dimensional Hilbert space per site. Going beyond finite-dimensional state space, recently Refs. [\onlinecite{DeMarcoPRL2021}, \onlinecite{demarco2021commuting}] proposed CPHs for chiral Abelian topological phases and U(1) SPT phases with non-zero Hall conductance in 2D. It is an interesting question to understand what phases can be represented by CPHs when infinite-dimensional local Hilbert spaces, such as quantum rotors, are allowed, and whether the stability results can be extended.

Our construction can be generalized to fermionic systems. A particularly interesting case is when the U(1) charge is the fermion number, which means that the charges decorated on vertices are fermionic. The anti-commutation relation between fermionic creation/annihilation operators means that the decoration can actually affect the local moves, leading to additional consistency conditions. We will describe the construction of fermionic topological insulators in an upcoming work. 

Our results in 3D cover two important classes of models: decorated Walker-Wang models to realize fractionally charged particles, and decorated Dijkgraaf-Witten models to realize U(1) fractionalization on loops. It will be interesting to find a more general construction that allows  both types of fractionalization to occur. 

\section{acknowledgement}
We are grateful to Yang Qi and Zheng-Cheng Gu for discussions and collaborations on a related project. We would like to thank Michael Levin for suggesting using cluster charge operator to compute fractional charge.
M. C. and Q. R. Wang are supported by NSF CAREER (DMR-1846109) and the Alfred P. Sloan foundation.

%

\appendix

\section{Hamiltonian of $\U$ symmetry-enriched $\Z_n$ toric code}
\label{App:H}

In this Appendix, we will present the explicit exactly-solvable CPH for the $\U$ symmetry-enriched $\Z_n$ toric code model.

\subsection{Definition of the Hamiltonian}

The Hamiltonian consists of two types of projection operator terms that commute with each other:
\begin{equation}\label{H_}
    H=-\sum_v A_v - \sum_p B_p.
\end{equation}
The vertex term $A_v$ enforces the fusion rule of the strings and decorates the $\U$ charge at the vertex $v$. It is a projection operator with expression
\begin{equation}\label{A}
    A_v = \delta_{[a_v+b_v-c_v],0}\delta_{S_v^z,\pm q(a_v,b_v)}
    =
    \begin{cases}
    1,&\mathrm{if}\ [a_v+b_v-c_v]=0\ \mathrm{and}\ S_v^z=\pm q(a_v,b_v)\\
    0,&\mathrm{otherwise}
    \end{cases}.
\end{equation}
Here, $a_v,b_v$ and $c_v$ are the group element labels of the three edges connecting vertex $v$. And the $\pm$ sign in front of $q(a_v,b_v)$ depends on the sublattice $A$ or $B$ of the vertex $v$. If and only if $A_v=1$, the string configurations and charge decoration at vertex $v$ satisfy the constraints shown in Eq.~(\ref{ZnV}).

To define a plaquette term $B_p$ that has nonzero action in the full string Hilbert space (including the string configurations with $[c]\ne [a+b]$ in the presence of $e$ anyons), we introduce a charge function $q_e$ as
\begin{equation}\label{q_AB}
    q_e(a,b,c)=\left\lfloor \frac{[a]+[b]-[c]}{n} \right\rfloor
    =
    \begin{cases}
    -1,&\mathrm{if}\ -(n-1)\le [a]+[b]-[c]<0\\
    0,&\mathrm{if}\ 0\le [a]+[b]-[c]<n\\
    1,&\mathrm{if}\ n\le [a]+[b]-[c]\le 2(n-1)
    \end{cases},
\end{equation}
for vertex $v$ of sublattice $A$ or $B$ with arbitrary string configurations $a,b,c\in\Z_n$:
\begin{equation}\label{ZnV_}
\vcenter{\hbox{
\begin{tikzpicture}[scale=0.25,>=stealth]
    \def\shift{0};
    \def\shifty{0}
	\draw[very thick] (\shift+0,\shifty+4) node[above]{$a$}  -- (\shift+2,\shifty+2) ;
	\draw[ very thick]  (\shift+4,\shifty+4) node[above]{$b$} -- (\shift+2,\shifty+2) ;
	\draw[ very thick](\shift+2,\shifty+2) -- (\shift+2,\shifty+0) node[below]{$c$};
    \filldraw (\shift+2, \shifty+2.1) circle (0.3) ;
	\node[right] at (\shift+2.2, \shifty+2.1) {$q_e(a,b,c)$};
    
    \def\shift{12};
    \def\shifty{0}
	\draw[ very thick] (\shift+0,\shifty+0) node[below]{$a$}  -- (\shift+2,\shifty+2) ;
	\draw[ very thick]  (\shift+4,\shifty+0) node[below]{$b$} -- (\shift+2,\shifty+2) ;
	\draw[ very thick](\shift+2,\shifty+2) -- (\shift+2,\shifty+4) node[above]{$c$};
    \filldraw (\shift+2, \shifty+1.9) circle (0.3);
    \node [right] at (\shift+2.2, \shifty+2.1) {$-q_e(a,b,c)$};
    \end{tikzpicture}
}}.
\end{equation}
Here, we use the floor function $\lfloor x\rfloor$ to denote the greatest integer less than or equal to $x$. If the string configuration satisfies the fusion rule $[c]=[a+b]$ in the ground state, the above charge function $q_e(a,b,a+b)$ reduces to the charge function $q(a,b)$ in Eq.~(\ref{q}):
\begin{equation}
    q_e(a,b,a+b)=q(a,b)=\frac{[a]+[b]-[a+b]}{n}.
\end{equation}
In this sense, the charge function $q_e(a,b,c)$ is a generalization of the $\U$ charge $q(a,b)$ from the ground state to the excited states in the presence of $e$ anyons.

The expression of the charge function $q_e(a,b,c)$ is obtained from a reference configuration using the global $\U$ charge conservation symmetry. We choose the neighborhood of an upper endpoint $v$ (of sublattice $A$) of an open string as the reference configuration [see the left figure of Eq.~(\ref{q_AB_})]. The $\U$ charge at $v$ is chosen to be $\delta_{[c-a-b]}-1$, i.e., $-1$ ($0$) if the open string label is nonzero (zero). Under the action of $B_{p_1}^{(a)}B_{p_2}^{(-b)}$, the string configuration near the vertex $v$ becomes the standard one that $a$ and $b$ fuse to $c$:
\begin{align}\label{q_AB_}
\vcenter{\hbox{
\begin{tikzpicture}[scale=0.8,>=stealth]
    \def\p1{1}
    \coordinate (p1) at (0,1);
    \coordinate (p2) at (0.8660,-.5);
    \coordinate (p3) at (-0.8660,-.5);
    \node[below right] at (0,0) {$v$};
    \foreach \p in {2,3} {
        \foreach \i in {0,1,2,3,4,5} \coordinate (p\p\i) at ($(p\p)+(90-60*\i:1)$);
    }
    \node[]at (p3) {$p_1$};
    \node[xshift=5]at (p2) {$p_2$};
    \foreach\i in {0,1,2,3,4,5} \draw[dotted,thick]($(p2)+(90-60*\i:1)$)--($(p2)+(90-60*\i+60:1)$);
    \foreach\i in {0,1,2,3,4,5} \draw[dotted,thick]($(p3)+(90-60*\i:1)$)--($(p3)+(90-60*\i+60:1)$);
    \draw[black,very thick](p23)--($(p23)+(0,-.4)$);
    \draw[black,very thick](p23)--(p24)--(p25);
    \draw[black,very thick]($(p20)+(0,.4)$)--(p20)--(p21)--(p22)--(p23);
    \draw[black,very thick]($(p30)+(0,.4)$)--(p30)--(p35)--(p34)--(p33)--(p32);
    \draw[dotted,thick]($(p33)+(0,-.4)$)--(p33);
    \filldraw (0,0) circle (0.1) node[yshift=8,scale=.8]{$0$ or $-$};
    \filldraw (p23) circle (0.1);
    \filldraw (p24) circle (0.1);
    \filldraw (p33) circle (0.1);
    \node[scale=.7,black,xshift=9,yshift=5]at ($(p23)!.5!(p24)$) {$c-b$};
    \node[scale=.7,black,xshift=-6,yshift=4]at ($(p33)!.5!(p32)$) {$-a$};
    \node[scale=.7,black,xshift=4]at ($(p25)!.5!(p24)$) {$c-a-b$};
    \node[scale=.7,black,below right]at (p23) {$c$};
    \node[scale=.7,black,above left]at (p30) {$a$};
    \node[scale=.7,black,above right]at (p20) {$b$};
    \node[scale=.7,black,right]at ($(p34)!.5!(p35)$) {$a$};
    \node[scale=.7,black,left]at ($(p21)!.5!(p22)$) {$b$};
\end{tikzpicture}
}}
\longrightarrow
\vcenter{\hbox{
\begin{tikzpicture}[scale=0.8,>=stealth]
    \def\p1{1}
    \coordinate (p1) at (0,1);
    \coordinate (p2) at (0.8660,-.5);
    \coordinate (p3) at (-0.8660,-.5);
    \node[below right] at (0,0) {$v$};
    \foreach \p in {2,3} {
        \foreach \i in {0,1,2,3,4,5} \coordinate (p\p\i) at ($(p\p)+(90-60*\i:1)$);
    }
    \node[]at (p3) {$p_1$};
    \node[]at (p2) {$p_2$};
    \foreach\i in {0,1,2,3,4,5} \draw[dotted,thick]($(p2)+(90-60*\i:1)$)--($(p2)+(90-60*\i+60:1)$);
    \foreach\i in {0,1,2,3,4,5} \draw[dotted,thick]($(p3)+(90-60*\i:1)$)--($(p3)+(90-60*\i+60:1)$);
    \draw[black,very thick](p23)--($(p23)+(0,-.4)$);
    \draw[black,very thick](p23)--(p24)--(p25);
    \draw[black,very thick]($(p20)+(0,.4)$)--(p20)--(p25);
    \draw[black,very thick]($(p30)+(0,.4)$)--(p30)--(p31);
    \draw[dotted,thick]($(p33)+(0,-.4)$)--(p33);
    \filldraw (0,0) circle (0.1) node[yshift=8,scale=.8]{$q_e$};
    \node[scale=.7,black,xshift=-5.5]at ($(p25)!.5!(p24)$) {$c$};
    \node[scale=.7,black,above left]at (p30) {$a$};
    \node[scale=.7,black,above right]at (p20) {$b$};
\end{tikzpicture}
}}.
\end{align}
Due to the global $\U$ charge conservation, the charge $q_e(a,b,c)$ of vertex $v$ of the right-hand-side state equals to the total charge of the left-hand-side state: $(\delta_{[c-a-b]}-1)+q(a,-a)+q(c-b,b)-q(-a,c-b)$. After some calculations, it is exactly $q_e(a,b,c)$ defined in Eq.~(\ref{q_AB}).

The plaquette term $B_p$ is designed to fluctuate the configurations of strings and charge decorations within the subspace with $S_v^z=\pm q_e(a_v,b_v,c_v)$ for every vertex $v$ (the $\pm$ sign again depends on the sublattice $A/B$). It is a summation of the terms $B_p^{(g)}$ labeled by $g\in\Z_n$ as
\begin{equation}
    B_p=\frac{1}{n}\sum_{g\in\Z_n}B_p^{(g)}.
\end{equation}
Each summand $B_p^{(g)}$ creates a small loop with label $g\in\Z_n$ in the plaquette $p$, and then fuses it to the six edges labeled by $g_i\in\Z_n$ ($1\le i\le 6$):
\begin{align}\label{q_AB_2}
\vcenter{\hbox{
\begin{tikzpicture}[scale=0.8,>=stealth]
    \def\p1{1}
    \coordinate (p1) at (0,0);
    \foreach \i in {0,1,2,3,4,5} \coordinate (p1\i) at ($(p1)+(90-60*\i:1)$);
    \foreach\i in {0,1,2,3,4,5} {
        \draw[black,very thick]($(p1)+(90-60*\i:1)$)--($(p1)+(90-60*\i+60:1)$);
        \draw[black,very thick](p1\i)--($(p1)+(90-60*\i:1.5)$);
        \filldraw (p1\i) circle (0.1);
    }
    \draw[dotted,thick] (p1) circle (.6);
    \draw [->,>=stealth] (.6,.1) -- (.6,.11);
    \node[scale=.7,black]at (.4,0) {$g$};
    \node[scale=.7,black,xshift=7,yshift=4]at (p10) {$v_1$};
    \node[scale=.7,black,above,yshift=1]at (p11) {$v_2$};
    \node[scale=.7,black,below,yshift=-1]at (p12) {$v_3$};
    \node[scale=.7,black,xshift=-7,yshift=-4]at (p13) {$v_4$};
    \node[scale=.7,black,below,yshift=-1]at (p14) {$v_5$};
    \node[scale=.7,black,above,yshift=1]at (p15) {$v_6$};
    \node[scale=.7,black,xshift=3,yshift=6]at ($(p10)!.5!(p11)$) {$g_1$};
    \node[scale=.7,black,xshift=7]at ($(p11)!.5!(p12)$) {$g_2$};
    \node[scale=.7,black,xshift=2,yshift=-6]at ($(p12)!.5!(p13)$) {$g_3$};
    \node[scale=.7,black,xshift=-2,yshift=-5]at ($(p13)!.5!(p14)$) {$g_4$};
    \node[scale=.7,black,xshift=-6]at ($(p14)!.5!(p15)$) {$g_5$};
    \node[scale=.7,black,xshift=-2,yshift=6]at ($(p10)!.5!(p15)$) {$g_6$};
    \node[scale=.7,black,xshift=6,yshift=0]at ($(p1)+(90-60*0:1.5)$) {$g_7$};
    \node[scale=.7,black,xshift=2,yshift=-6]at ($(p1)+(90-60*1:1.5)$) {$g_8$};
    \node[scale=.7,black,xshift=-1,yshift=-5]at ($(p1)+(90-60*2:1.5)$) {$g_9$};
    \node[scale=.7,black,xshift=8,yshift=0]at ($(p1)+(90-60*3:1.5)$) {$g_{10}$};
    \node[scale=.7,black,xshift=1,yshift=-5]at ($(p1)+(90-60*4:1.5)$) {$g_{11}$};
    \node[scale=.7,black,xshift=-2,yshift=-6]at ($(p1)+(90-60*5:1.5)$) {$g_{12}$};
\end{tikzpicture}
}}
\xrightarrow[]{B_p^{(g)}}
\vcenter{\hbox{
\begin{tikzpicture}[scale=0.8,>=stealth]
    \def\p1{1}
    \coordinate (p1) at (0,0);
    \foreach \i in {0,1,2,3,4,5} \coordinate (p1\i) at ($(p1)+(90-60*\i:1)$);
    \foreach\i in {0,1,2,3,4,5} {
        \draw[black,very thick]($(p1)+(90-60*\i:1)$)--($(p1)+(90-60*\i+60:1)$);
        \draw[black,very thick](p1\i)--($(p1)+(90-60*\i:1.5)$);
        \filldraw (p1\i) circle (0.1);
    }
    \node[scale=.7,black,xshift=7,yshift=4]at (p10) {$v_1$};
    \node[scale=.7,black,above,yshift=1]at (p11) {$v_2$};
    \node[scale=.7,black,below,yshift=-1]at (p12) {$v_3$};
    \node[scale=.7,black,xshift=-7,yshift=-4]at (p13) {$v_4$};
    \node[scale=.7,black,below,yshift=-1]at (p14) {$v_5$};
    \node[scale=.7,black,above,yshift=1]at (p15) {$v_6$};
    \node[scale=.7,black,xshift=4,yshift=7]at ($(p10)!.5!(p11)$) {$g_1'$};
    \node[scale=.7,black,xshift=7]at ($(p11)!.5!(p12)$) {$g_2'$};
    \node[scale=.7,black,xshift=2,yshift=-5]at ($(p12)!.5!(p13)$) {$g_3'$};
    \node[scale=.7,black,xshift=-4,yshift=-6]at ($(p13)!.5!(p14)$) {$g_4'$};
    \node[scale=.7,black,xshift=-6]at ($(p14)!.5!(p15)$) {$g_5'$};
    \node[scale=.7,black,xshift=-2,yshift=7]at ($(p10)!.5!(p15)$) {$g_6'$};
    \node[scale=.7,black,xshift=6,yshift=0]at ($(p1)+(90-60*0:1.5)$) {$g_7$};
    \node[scale=.7,black,xshift=2,yshift=-6]at ($(p1)+(90-60*1:1.5)$) {$g_8$};
    \node[scale=.7,black,xshift=-1,yshift=-5]at ($(p1)+(90-60*2:1.5)$) {$g_9$};
    \node[scale=.7,black,xshift=8,yshift=0]at ($(p1)+(90-60*3:1.5)$) {$g_{10}$};
    \node[scale=.7,black,xshift=1,yshift=-5]at ($(p1)+(90-60*4:1.5)$) {$g_{11}$};
    \node[scale=.7,black,xshift=-2,yshift=-6]at ($(p1)+(90-60*5:1.5)$) {$g_{12}$};
\end{tikzpicture}
}}.
\end{align}
The new group element labels of the six edges become $g_i'=[g_i+g]$ ($1\le i\le 3$) and $g_i'=[g_i-g]$ ($4\le i\le 6$). And all outer edges of the plaquette are unchanged: $g_i'=g_i$ ($7\le i\le 12$). At the same time, the $\U$ charges of each vertex $v$ before and after the action of $B_p^{(g)}$ are simply projected to the subspace with $S_v^z=\pm q_{e}(a_v,b_v,c_v)$ and $S_v^z=\pm q_{e}(a_v',b_v',c_v')$, respectively. Formally, the operator $B_p^{(g)}$ can be written as
\begin{equation}\label{Bpg}
    B_p^{(g)}=\left(Z_{g_1}Z_{g_2}Z_{g_3}\right)^g \left(Z_{g_4}Z_{g_5}Z_{g_6}\right)^{-g} \cdot W_{v_1}W_{v_2}W_{v_3}W_{v_4}W_{v_5}W_{v_6}\cdot \prod_v P_v.
\end{equation}
The first term $Z_{g_i}$ is the shift matrix of the edge label $g_i$ as
\begin{equation}
    Z_{g_i}=\sum_{g_i\in\Z_n}|[g_i+1]\rangle \langle g_i|.
\end{equation}
The last term $P_v$ for vertex $v$ of sublattice $A/B$ is a projection operator to the subspace with correct charge decorations:
\begin{equation}
    P_v=\delta_{S_v^z,\pm q_{e}(a_v,b_v,c_v)},
\end{equation}
where $a_v,b_v$ and $c_v$ are the three edge labels of vertex $v$.
With this projection operator, $B_p^{(g)}$ is nonzero only if the initial configuration satisfies the charge decoration condition. In this subspace, the operator $W_v$ changes the $\U$ charge decorations from the initial string configuration $\{g_i\}$ to the final one $\{g_i'\}$. The formal expression of $W_v$ is
\begin{equation}
    W_{v}=|S_v^z=\pm q_{e}(a_v',b_v',c_v')\rangle \langle S_v^z=\pm q_{e}(a_v,b_v,c_v)|.
\end{equation}
With all these terms, $B_p$ fluctuate all the possible string configurations of $\Z_n$ toric code with $\U$ charge decorations given by $S_v^z=\pm q_{e}(a_v,b_v,c_v)$ for all vertex $v$.

The ground state of this model is a simultaneous eigenstate of $A_v$ and $B_p$ with eigenvalue $1$. In terms of basis states, it is a equal-weight superposition of all possible string configurations with $[c_v]=[a_v+b_v]$ and $S_v^z=\pm q_e(a_v,b_v,a_v+b_v) = \pm q(a_v,b_v)$ for all vertex $v$.

\subsection{Algebraic relations of $A_v$ and $B_p$}

According to the definition Eq.~(\ref{A}) which is diagonal in the configuration basis states, it is easy to show that $A_v$ satisfies the relations
\begin{align}
    A_v^\dagger &=A_v,\\
    (A_v)^2&=A_v,\\
    [A_v,A_{v'}]&=0.
\end{align}
So they are Hermitian projection operators that commute with each other.

For the plaquette operator $B_p^{(g)}$, one can also show that
\begin{align}
    \left(B_p^{(g)}\right)^\dagger &= B_p^{([-g])},\\
    B_p^{(g)} B_p^{(h)} &= B_p^{([g+h])},\label{BB}\\
    [B_p^{(g)},B_{p'}^{(h)}]&=0,
\end{align}
for arbitrary plaquettes $p,p'$ and group elements $g,h\in\Z_n$. Using these relations, we have
\begin{align}
    B_p^\dagger &= B_p,\\
    (B_p)^2&=B_p,\\
    [B_p,B_{p'}]&=0.
\end{align}
So $B_p$'s are also Hermitian projection operators that commute with each other.

It is also easy to show that $B_p$ does not mix the states in the subspace of $A_v=1$ and $A_v=0$. So they also commute with each other:
\begin{equation}
    [A_v,B_{p}]=0.
\end{equation}
With all the above algebra relations, the Hamiltonian we constructed is an exactly solvable CPH.

\subsection{$\U$ charge conservation}

In this section, we will show that the Hamiltonian preserves the global $\U$ charge symmetry. The operator $A_v$ is already diagonal in the charge basis, so we only need to check that $B_p^{(g)}$ preserves the $\U$ charge. Using the relation Eq.~(\ref{BB}), the problem is further reduced to the $\U$ charge conservation of $B_p^{(g=1)}$.

Before the action of $B_p^{(g=1)}$, the total charge of the left-hand configuration of Eq.~(\ref{q_AB_2}) is
\begin{equation}
    Q=q_e(g_1,g_8,g_2)
    +q_e(g_4,g_3,g_{10})
    +q_e(g_{12},g_{6},g_5)
    -q_e(g_6,g_1,g_7)
    -q_e(g_3,g_9,g_2)
    -q_e(g_{11},g_{4},g_5).
\end{equation}
After the action, the total charge of the right-hand configuration of Eq.~(\ref{q_AB_2}) is
\begin{align}\nonumber
    Q'&=q_e(g_1+1,g_8,g_2+1)+q_e(g_4-1,g_3+1,g_{10})-q_e(g_{12},g_6-1,g_5-1)\\
    &\quad-q_e(g_6-1,g_1+1,g_7)-q_e(g_3+1,g_9,g_2+1)-q_e(g_{11},g_4-1,g_5-1).
\end{align}
It is not hard to show that the charge function $q_e(a,b,c)$ defined in Eq.~(\ref{q_AB}) satisfies
\begin{align}
    q_e(a+1,b,c+1)-q_e(a,b,c) &= \delta_{[c],n-1}-\delta_{[a],n-1},\\
    q_e(a,b-1,c-1)-q_e(a,b,c) &= \delta_{[b],0}-\delta_{[c],0},\\
    q_e(a-1,b+1,c)-q_e(a,b,c) &= \delta_{[a],0}-\delta_{[b],n-1}.
\end{align}
Using these relations, the charge difference between the initial and final configurations under the action of $B_p^{(g=1)}$ is
\begin{align}\nonumber
    Q'-Q&=
    \left(\delta_{[g_2],n-1}-\delta_{[g_1],n-1}\right)
    +\left(\delta_{[g_4],0}-\delta_{[g_3],n-1}\right)
    +\left(\delta_{[g_6],0}-\delta_{[g_5],0}\right)\\\nonumber
    &\quad-\left(\delta_{[g_6],0}-\delta_{[g_1],n-1}\right)
    -\left(\delta_{[g_2],n-1}-\delta_{[g_3],n-1}\right)
    -\left(\delta_{[g_4],0}-\delta_{[g_5],0}\right)\\
    &=0.
\end{align}
As a result, we conclude that $B_p^{(g=1)}$, and hence the total Hamiltonian, preserve the global $\U$ charge symmetry.

\section{Fractional charge of $e$ anyons}
\label{App:frac}

In this appendix, we will illustrate the full action of $B_{p_3}B_{p_2}B_{p_1}$ in Eq.~(\ref{Psi_e}) for a larger lattice patch to show the symmetry fractionalization of $e$ anyons in $\Z_2$ toric code.

Similar to Eq.~(\ref{Psi_e}) in the main text, let us assume that $v$ is the upper endpoint of an open string in the initial state. The initial $\U$ charge of $v$ is chosen to be $Q_v=-1$. A string configuration will split into two under the action of the plaquette operator $B_{p} = \frac{1}{2}(1+B_{p}^{s})$, where $B_{p}^{s}$ creates a small loop of string $s$ around the plaquette $p$. The details of the plaquette operator is defined in Eq.~(\ref{Bpg}) with $n=2$.
Under the action of three plaquette terms $B_{p_3}B_{p_2}B_{p_1}$ around the vertex $v$, the initial state is turned to eight different configurations:
\begin{align}\nonumber\label{Bp321}
B_{p_3}B_{p_2}B_{p_1}
\vcenter{\hbox{
\begin{tikzpicture}[scale=0.4,>=stealth]
    \def\p1{1}
    \coordinate (p1) at (0,1);
    \coordinate (p2) at (0.8660,-.5);
    \coordinate (p3) at (-0.8660,-.5);
    \foreach \p in {1,2,3} {
        \node[]at (p\p) {$p_\p$};
        \foreach \i in {0,1,2,3,4,5} \coordinate (p\p\i) at ($(p\p)+(90-60*\i:1)$);
    }
    \foreach\i in {0,1,2,3,4,5} \draw[dotted,thick]($(p1)+(90-60*\i:1)$)--($(p1)+(90-60*\i+60:1)$);
    \foreach\i in {1,2,3,4,5} \draw[dotted,thick]($(p2)+(90-60*\i:1)$)--($(p2)+(90-60*\i+60:1)$);
    \foreach\i in {3,4,5,6} \draw[dotted,thick]($(p3)+(90-60*\i:1)$)--($(p3)+(90-60*\i+60:1)$);
    \draw[blue,thick](p23)--($(p23)+(0,-.3)$);
    \draw[blue,thick](p23)--(p24)--(p25);
    \filldraw (0,0) circle (0.15) node[above,scale=.7]{$-$};
\end{tikzpicture}
}}
&=
\frac{1}{8}
\left(
\vcenter{\hbox{
\begin{tikzpicture}[scale=0.4,>=stealth]
    \def\p1{1}
    \coordinate (p1) at (0,1);
    \coordinate (p2) at (0.8660,-.5);
    \coordinate (p3) at (-0.8660,-.5);
    \foreach \p in {1,2,3} {
        \node[]at (p\p) {$p_\p$};
        \foreach \i in {0,1,2,3,4,5} \coordinate (p\p\i) at ($(p\p)+(90-60*\i:1)$);
    }
    \foreach\i in {0,1,2,3,4,5} \draw[dotted,thick]($(p1)+(90-60*\i:1)$)--($(p1)+(90-60*\i+60:1)$);
    \foreach\i in {1,2,3,4,5} \draw[dotted,thick]($(p2)+(90-60*\i:1)$)--($(p2)+(90-60*\i+60:1)$);
    \foreach\i in {3,4,5,6} \draw[dotted,thick]($(p3)+(90-60*\i:1)$)--($(p3)+(90-60*\i+60:1)$);
    \draw[blue,thick](p23)--($(p23)+(0,-.3)$);
    \draw[blue,thick](p23)--(p24)--(p25);
    \filldraw (0,0) circle (0.15) node[above,scale=.7]{$-$};
\end{tikzpicture}
}}
+
\vcenter{\hbox{
\begin{tikzpicture}[scale=0.4,>=stealth]
    \def\p1{1}
    \coordinate (p1) at (0,1);
    \coordinate (p2) at (0.8660,-.5);
    \coordinate (p3) at (-0.8660,-.5);
    \foreach \p in {1,2,3} {
        \node[]at (p\p) {$p_\p$};
        \foreach \i in {0,1,2,3,4,5} \coordinate (p\p\i) at ($(p\p)+(90-60*\i:1)$);
    }
    \foreach\i in {0,1,2,3,4,5} \draw[dotted,thick]($(p1)+(90-60*\i:1)$)--($(p1)+(90-60*\i+60:1)$);
    \foreach\i in {1,2,3,4,5} \draw[dotted,thick]($(p2)+(90-60*\i:1)$)--($(p2)+(90-60*\i+60:1)$);
    \foreach\i in {3,4,5,6} \draw[dotted,thick]($(p3)+(90-60*\i:1)$)--($(p3)+(90-60*\i+60:1)$);
    \draw[blue,thick](p23)--($(p23)+(0,-.3)$);
    \draw[blue,thick](p23)--(p24)--(p25);
    \draw[blue,thick](p10)--(p11)--(p12)--(p13)--(p14)--(p15)--cycle;
    \filldraw (p10) circle (0.15) node[above,scale=.7]{$-$};
\end{tikzpicture}
}}
+
\vcenter{\hbox{
\begin{tikzpicture}[scale=0.4,>=stealth]
    \def\p1{1}
    \coordinate (p1) at (0,1);
    \coordinate (p2) at (0.8660,-.5);
    \coordinate (p3) at (-0.8660,-.5);
    \foreach \p in {1,2,3} {
        \node[]at (p\p) {$p_\p$};
        \foreach \i in {0,1,2,3,4,5} \coordinate (p\p\i) at ($(p\p)+(90-60*\i:1)$);
    }
    \foreach\i in {0,1,2,3,4,5} \draw[dotted,thick]($(p1)+(90-60*\i:1)$)--($(p1)+(90-60*\i+60:1)$);
    \foreach\i in {1,2,3,4,5} \draw[dotted,thick]($(p2)+(90-60*\i:1)$)--($(p2)+(90-60*\i+60:1)$);
    \foreach\i in {3,4,5,6} \draw[dotted,thick]($(p3)+(90-60*\i:1)$)--($(p3)+(90-60*\i+60:1)$);
    \draw[blue,thick](p23)--($(p23)+(0,-.3)$);
    \draw[blue,thick](p23)--(p22)--(p21)--(p20)--(p25);
    \filldraw (p20) circle (0.15) node[below,scale=.7]{$-$};
\end{tikzpicture}
}}
+
\vcenter{\hbox{
\begin{tikzpicture}[scale=0.4,>=stealth]
    \def\p1{1}
    \coordinate (p1) at (0,1);
    \coordinate (p2) at (0.8660,-.5);
    \coordinate (p3) at (-0.8660,-.5);
    \foreach \p in {1,2,3} {
        \node[]at (p\p) {$p_\p$};
        \foreach \i in {0,1,2,3,4,5} \coordinate (p\p\i) at ($(p\p)+(90-60*\i:1)$);
    }
    \foreach\i in {0,1,2,3,4,5} \draw[dotted,thick]($(p1)+(90-60*\i:1)$)--($(p1)+(90-60*\i+60:1)$);
    \foreach\i in {1,2,3,4,5} \draw[dotted,thick]($(p2)+(90-60*\i:1)$)--($(p2)+(90-60*\i+60:1)$);
    \foreach\i in {3,4,5,6} \draw[dotted,thick]($(p3)+(90-60*\i:1)$)--($(p3)+(90-60*\i+60:1)$);
    \draw[blue,thick](p23)--($(p23)+(0,-.3)$);
    \draw[blue,thick](p23)--(p22)--(p21)--(p20);
    \draw[blue,thick](p13)--(p14)--(p15)--(p10)--(p11)--(p12);
    \filldraw (p10) circle (0.15) node[above,scale=.7]{$-$};
\end{tikzpicture}
}}
\right)\\
&\quad+
\frac{1}{8}
\left(
\vcenter{\hbox{
\begin{tikzpicture}[scale=0.4,>=stealth]
    \def\p1{1}
    \coordinate (p1) at (0,1);
    \coordinate (p2) at (0.8660,-.5);
    \coordinate (p3) at (-0.8660,-.5);
    \foreach \p in {1,2,3} {
        \node[]at (p\p) {$p_\p$};
        \foreach \i in {0,1,2,3,4,5} \coordinate (p\p\i) at ($(p\p)+(90-60*\i:1)$);
    }
    \foreach\i in {0,1,2,3,4,5} \draw[dotted,thick]($(p1)+(90-60*\i:1)$)--($(p1)+(90-60*\i+60:1)$);
    \foreach\i in {1,2,3,4,5} \draw[dotted,thick]($(p2)+(90-60*\i:1)$)--($(p2)+(90-60*\i+60:1)$);
    \foreach\i in {3,4,5,6} \draw[dotted,thick]($(p3)+(90-60*\i:1)$)--($(p3)+(90-60*\i+60:1)$);
    \draw[blue,thick](p23)--($(p23)+(0,-.3)$);
    \draw[blue,thick](p23)--(p24);
    \draw[blue,thick](p32)--(p33)--(p34)--(p35)--(p30)--(p31);
    \filldraw (p30) circle (0.15) node[below,scale=.7]{$-$};
    \filldraw (p33) circle (0.15) node[above,scale=.7]{$+$};
    \filldraw (p32) circle (0.15) node[below,scale=.7]{$-$};
\end{tikzpicture}
}}
+
\vcenter{\hbox{
\begin{tikzpicture}[scale=0.4,>=stealth]
    \def\p1{1}
    \coordinate (p1) at (0,1);
    \coordinate (p2) at (0.8660,-.5);
    \coordinate (p3) at (-0.8660,-.5);
    \foreach \p in {1,2,3} {
        \node[]at (p\p) {$p_\p$};
        \foreach \i in {0,1,2,3,4,5} \coordinate (p\p\i) at ($(p\p)+(90-60*\i:1)$);
    }
    \foreach\i in {0,1,2,3,4,5} \draw[dotted,thick]($(p1)+(90-60*\i:1)$)--($(p1)+(90-60*\i+60:1)$);
    \foreach\i in {1,2,3,4,5} \draw[dotted,thick]($(p2)+(90-60*\i:1)$)--($(p2)+(90-60*\i+60:1)$);
    \foreach\i in {3,4,5,6} \draw[dotted,thick]($(p3)+(90-60*\i:1)$)--($(p3)+(90-60*\i+60:1)$);
    \draw[blue,thick](p23)--($(p23)+(0,-.3)$);
    \draw[blue,thick](p23)--(p24)--(p33)--(p34)--(p35)--(p30);
    \draw[blue,thick](p14)--(p15)--(p10)--(p11)--(p12)--(p13);
    \filldraw (p10) circle (0.15) node[above,scale=.7]{$-$};
    \filldraw (p32) circle (0.15) node[below,scale=.7]{$-$};
    \filldraw (p33) circle (0.15) node[above,scale=.7]{$+$};
\end{tikzpicture}
}}
+
\vcenter{\hbox{
\begin{tikzpicture}[scale=0.4,>=stealth]
    \def\p1{1}
    \coordinate (p1) at (0,1);
    \coordinate (p2) at (0.8660,-.5);
    \coordinate (p3) at (-0.8660,-.5);
    \foreach \p in {1,2,3} {
        \node[]at (p\p) {$p_\p$};
        \foreach \i in {0,1,2,3,4,5} \coordinate (p\p\i) at ($(p\p)+(90-60*\i:1)$);
    }
    \foreach\i in {0,1,2,3,4,5} \draw[dotted,thick]($(p1)+(90-60*\i:1)$)--($(p1)+(90-60*\i+60:1)$);
    \foreach\i in {1,2,3,4,5} \draw[dotted,thick]($(p2)+(90-60*\i:1)$)--($(p2)+(90-60*\i+60:1)$);
    \foreach\i in {3,4,5,6} \draw[dotted,thick]($(p3)+(90-60*\i:1)$)--($(p3)+(90-60*\i+60:1)$);
    \draw[blue,thick](p23)--($(p23)+(0,-.3)$);
    \draw[blue,thick](p23)--(p22)--(p21)--(p20)--(p25);
    \draw[blue,thick](p30)--(p31)--(p32)--(p33)--(p34)--(p35)--(p30);
    \filldraw (p20) circle (0.15) node[below,scale=.7]{$-$};
    \filldraw (p30) circle (0.15) node[below,scale=.7]{$-$};
    \filldraw (p33) circle (0.15) node[above,scale=.7]{$+$};
\end{tikzpicture}
}}
+
\vcenter{\hbox{
\begin{tikzpicture}[scale=0.4,>=stealth]
    \def\p1{1}
    \coordinate (p1) at (0,1);
    \coordinate (p2) at (0.8660,-.5);
    \coordinate (p3) at (-0.8660,-.5);
    \foreach \p in {1,2,3} {
        \node[]at (p\p) {$p_\p$};
        \foreach \i in {0,1,2,3,4,5} \coordinate (p\p\i) at ($(p\p)+(90-60*\i:1)$);
    }
    \foreach\i in {0,1,2,3,4,5} \draw[dotted,thick]($(p1)+(90-60*\i:1)$)--($(p1)+(90-60*\i+60:1)$);
    \foreach\i in {1,2,3,4,5} \draw[dotted,thick]($(p2)+(90-60*\i:1)$)--($(p2)+(90-60*\i+60:1)$);
    \foreach\i in {3,4,5,6} \draw[dotted,thick]($(p3)+(90-60*\i:1)$)--($(p3)+(90-60*\i+60:1)$);
    \draw[blue,thick](p23)--($(p23)+(0,-.3)$);
    \draw[blue,thick](p23)--(p22)--(p21)--(p20);
    \draw[blue,thick](p14)--(p15)--(p10)--(p11)--(p12);
    \draw[blue,thick](p31)--(p32)--(p33)--(p34)--(p35)--(p30);
    \filldraw (p33) circle (0.15) node[above,scale=.7]{$+$};
    \filldraw (p10) circle (0.15) node[above,scale=.7]{$-$};
    \filldraw (0,0) circle (0.15) node[above,scale=.7]{$-$};
\end{tikzpicture}
}}
\right).
\end{align}
This is a state for larger lattice compared to the hand-waving pictures in Eq.~(\ref{Psi_e}). Each local patch of the eight configurations has a total $\U$ charge $-1$, which is the same as the initial state due to charge conservation.

Locally, the $e$ anyon excitation has surrounding string configurations and charge decorations as Eq.~(\ref{Bp321}) in the excited states. So the average charge of vertex $v$ in the excited state is
\begin{equation}
    \langle Q_v\rangle = -2/8 = -1/4,
\end{equation}
which matches with the argument below Eq.~(\ref{Psi_e}) in the main text. On the other hand, for a vertex different from the string endpoints, the local string configurations and charge decorations of the excited state are exactly the same as in the ground state. So their average charges are the same as that of the vacuum [see the discussion below Eq.~(\ref{Psi_0})]. In summary, the $e$ anyon has half $\U$ charge compared to the ground state background.

\end{document}